\documentclass[conference]{IEEEtran}

\let\labelindent\relax
\usepackage{enumitem}
\usepackage{nicefrac}
\usepackage{siunitx}
\usepackage{arydshln}
\usepackage{array,framed}
\usepackage{cite}
\usepackage{booktabs}
\usepackage{bm}

\usepackage{
  color,
  float,
  epsfig,
  wrapfig,
  graphics,
  graphicx,
  subcaption
}
\usepackage{textcomp,amssymb}
\usepackage{setspace}

\usepackage{caption}
\usepackage{nameref}
\usepackage{latexsym,fancyhdr,url}
\usepackage{enumerate}
\usepackage{algpseudocode}
\usepackage{graphics}
\usepackage{adjustbox}
\usepackage{xparse} 
\usepackage{xspace}
\usepackage{multirow}
\usepackage{csvsimple}
\usepackage{balance}

\usepackage{
  tikz,
  pgfplots,
  pgfplotstable
}

\usepackage{amsthm}
\newtheorem{theorem}{Theorem}
\usepackage[most]{tcolorbox}
\usepackage{xcolor}   
\usepackage{hyperref}
\hypersetup{                    
  colorlinks,
  linkcolor={green!80!black},
  citecolor={red!70!black},
  urlcolor={blue!70!black}
}
\definecolor{takeaways}{HTML}{F5F5F5}

\usetikzlibrary{
  shapes.geometric,
  arrows,
  external,
  pgfplots.groupplots,
  matrix
}

\pgfplotsset{compat=1.9}

\usepackage{mathtools}

\DeclareMathAlphabet{\mathcal}{OMS}{cmsy}{m}{n}

\DeclareGraphicsExtensions{%
    .png,.PNG,%
    .pdf,.PDF,%
    .jpg,.mps,.jpeg,.jbig2,.jb2,.JPG,.JPEG,.JBIG2,.JB2}

\usepackage{bbm}
\definecolor{ref}{HTML}{116D6E}
\definecolor{cite}{HTML}{E55807}
\definecolor{takeaways}{HTML}{F5F5F5}
\DeclareMathAlphabet{\mathbf}{OMS}{cmsy}{m}{n}
\hypersetup{                    
  colorlinks,
  linkcolor={green!80!black},
  citecolor={red!70!black},
  urlcolor={blue!70!black}
}
\usepackage{xurl}

\usepackage{amsthm}
\usepackage{multirow}
\usepackage{pifont}
\usepackage{caption}
\usepackage{graphicx}  
\usepackage{subcaption} 
\usepackage{algorithm}
\usepackage{algpseudocode}
\allowdisplaybreaks

\renewcommand{\paragraph}[1]{\vspace*{6pt}\noindent\textbf{#1}\;}
\newcounter{mypar}
\newcommand{\myparagraph}[1]{%
  \refstepcounter{mypar}%
  \paragraph{Attack-\Roman{mypar} #1}\label{Attack-\Roman{mypar}}%
}

\newcommand{\blackball}[1][0.15cm]{
    \tikz{\fill[black] (0,0) circle (#1);}
}

\newcommand{\whiteball}[1][0.15cm]{
    \tikz{\fill[white] (0,0) circle (#1); \draw (0,0) circle (#1);}
}

\newcommand{\halfball}[1][0.15cm]{
    \tikz{
        \fill[black] (0,0) -- (0,-#1) arc (-90:90:#1) -- cycle;
        \fill[white] (0,0) -- (0,#1) arc (90:270:#1) -- cycle;
        \draw (0,0) circle (#1);
    }
}

\hypersetup{
  colorlinks,
  linkcolor={green!80!black},
  citecolor={red!70!black},
  urlcolor={blue!70!black}
}

\renewcommand{\Pr}[1]{\ensuremath{\mathsf{Pr}\left[#1\right]}\xspace}

\begin{document}
%
\title{Black-box Membership Inference Attacks against Fine-tuned Diffusion Models}

\author{\IEEEauthorblockN{Yan Pang}
\IEEEauthorblockA{University of Virginia\\
trv3px@virginia.edu}
\and
\IEEEauthorblockN{Tianhao Wang}
\IEEEauthorblockA{University of Virginia\\
tianhao@virginia.edu}}

\IEEEoverridecommandlockouts
\makeatletter\def\@IEEEpubidpullup{6.5\baselineskip}\makeatother
\IEEEpubid{\parbox{\columnwidth}{
		Network and Distributed System Security (NDSS) Symposium 2025\\
		23 - 28 February 2025, San Diego, CA, USA\\
		ISBN 979-8-9894372-8-3\\
		https://dx.doi.org/10.14722/ndss.2025.23324\\
		www.ndss-symposium.org
}
\hspace{\columnsep}\makebox[\columnwidth]{}}

\maketitle

\begin{abstract}
With the rapid advancement of diffusion-based image-generative models, the quality of generated images has become increasingly photorealistic. Moreover, with the release of high-quality pre-trained image-generative models, a growing number of users are downloading these pre-trained models to fine-tune them with downstream datasets for various image-generation tasks. However, employing such powerful pre-trained models in downstream tasks presents significant privacy leakage risks. In this paper, we propose the first scores-based membership inference attack framework~\footnote{Code accessible at \url{https://github.com/py85252876/Reconstruction-based-Attack}} tailored for recent diffusion models, and in the more stringent black-box access setting. Considering four distinct attack scenarios and three types of attacks, this framework is capable of targeting any popular conditional generator model, achieving high precision, evidenced by an impressive AUC of $0.95$.
\end{abstract}


%

\section{Introduction}
The recent developments in image-generative models have been remarkably swift, and many applications based on these models have appeared.
Diffusion models~\cite{ramesh2021zero,ho2020denoising,nichol2022glide,ramesh2022hierarchical,saharia2022photorealistic,sohl2015deep,song2021scorebased,rombach2022high,ho2022classifierfree,song2022denoising} have come to the forefront of image generation. These models generate target images by progressive denoising a noisy sample from an isotropic Gaussian distribution. 
In an effort to expedite the training of diffusion models and reduce training expenses, Stable Diffusion~\cite{rombach2022high} was introduced. Leveraging the extensive and high-fidelity LAION-2B~\cite{schuhmann2022laion5b} dataset for training, the Stable Diffusion pre-trained checkpoint, available on HuggingFace, can be fine-tuned efficiently with just a few steps for effective deployment in downstream tasks. This model's efficiency has spurred an increasing number of usages of Stable Diffusion.

{At the same time, there has been a significant amount of research focused on the privacy concerns associated with these models, specifically those related to training data~\cite{duan2023diffusion, hu2023membership, wu2022membership,pang2023white,li2023meticulously} and those related to model outputs~\cite{shan2023glaze,chou2024villandiffusion,peng2023protecting}.} Among them, membership inference attacks (MIAs) primarily investigate whether a given sample $x$ is included in the training set of a specific target model. While this line of research was traditionally directed toward classifier models~\cite{choquette2021label,hui2021practical,rezaei2021difficulty,sablayrolles2019whitebox,salem2018mlleaks,shokri2017membership,song2021systematic,truex2019demystifying,yeom2018privacy}, the popularity of diffusion models has led to the application of MIAs to examine potential abuses of privacy in their training datasets. Depending on the level of access to the target model, these attacks can be categorized into white-box attacks, gray-box attacks, and black-box attacks. 

In a white-box attack scenario, attackers have access to all parameters of a model. Similar to membership inference attack targeting classifiers, attacks against diffusion models also utilize internal model information such as loss~\cite{carlini2023extracting,hu2023membership,matsumoto2023membership} or gradients~\cite{pang2023white} as attack features. 
Hu et al.~\cite{hu2023membership} and Matsumoto et al.~\cite{matsumoto2023membership} have utilized losses at different timesteps of querying the model as attack features. Similarly, Carlini et al.~\cite{carlini2023extracting} employed losses across various timesteps but incorporated the LiRA framework to construct two distributions for inferring the membership of a sample $x$. Pang et al.~\cite{pang2023white} took a different approach by using the model's gradients at different timesteps as the attack features, positing that gradient information better reflects the model's response to $x$.

Although white-box attacks can achieve high success rates, their limitation lies in the requirement for complete access to the target model's information, which is often impractical in real-world scenarios. Compared with white-box attack, gray-box approaches do not require full access to the model's parameters; instead, they only necessitate the intermediate outputs from the diffusion model during the denoising process to serve as features for inference~\cite{duan2023diffusion,fu2023probabilistic,hu2023membership,kong2023efficient,li2024unveiling,fu2024model,zhai2024membership}. 
For example, Duan et al.~\cite{duan2023diffusion}, and Kong et al.~\cite{kong2023efficient} have leveraged the deterministic nature of DDIMs, using the approximated posterior estimation error of intermediate outputs at different timesteps as attack features. Hu et al.~\cite{hu2023membership} have proposed using intermediate outputs to estimate the log-likelihood of samples as attack features. However, these attacks inevitably rely on the intermediate images generated during the model's operation. In real-world scenarios, if a malicious model is trained using private or unsafe images, typically only the final output image is provided, with efforts made to conceal as many model details as possible. Therefore, the more practical scenarios would be black-box.

There are also black-box attacks for GANs~\cite{chen2020gan, hilprecht2019monte} and VAEs~\cite{hilprecht2019monte}. These are based on {\it unconditional} generative models and involve a highly stochastic generation process that requires {\it extensive sampling} for inference, which becomes inefficient when directly applied to diffusion models. The other black-box attacks~\cite{matsumoto2023membership,wu2022membership,zhang2024generated, dubinski2024towards}, although more tailored for diffusion models, focus on simulations and lack the necessary conditions to be used in realistic scenarios. We will discuss them in \autoref{subsec:existing}.

%
\begin{figure*}
    \centering
    \includegraphics[width=0.98\linewidth]{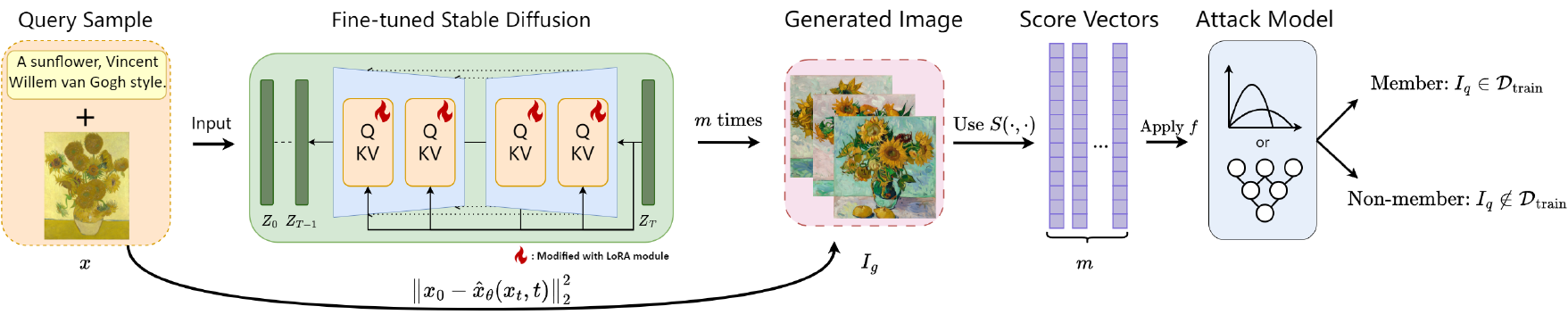}
    \caption{{Our attack takes the query sample $x$, which consists of an image $I_q$ and a text component $T_q$, and applies $T_q$ to query the model to get generated image $I_g$ for $m$ times. Then, we compute the similarity score between $I_q$ and each $I_g$ with $S(\cdot,\cdot)$. The $m$ scores are then aggregated using $f$, and used to train the attack model to determine the membership.}}
    \label{fig:overview}
\end{figure*}

In this paper, we present a black-box attack framework suitable for state-of-the-art image generative models, {as shown in~\autoref{fig:overview}}.
The framework was built on a careful analysis of the objective function of
the diffusion model as its theoretical foundation, and compares the generated image and the query image. It also incorporates four potential attack scenarios tailored for different settings of diffusion models.
We demonstrate the efficacy of our attack using the pre-trained Stable Diffusion v1-5 and further validate it fine-tuned with CelebA-Dialog~\cite{jiang2021talkedit}, WIT~\cite{srinivasan2021wit}, and MS COCO datasets~\cite{lin2015microsoft}.

Compared with existing black-box methods~\cite{matsumoto2023membership,zhang2024generated}, our attack under four attack scenarios can achieve $87\%$ accuracy and outperform other methods by nearly $35\%$. We systematically evaluate all components, including the image encoder, distance metrics, inference steps, and training set sizes. Our method is able to achieve high ROC-AUC across three datasets: $0.95$, $0.85$, and $0.93$. Even using different {types of generative models} as shadow models to employ the attack, our attack still can obtain at least $83\%$ success rate for four attack scenarios on three datasets. The results show that our attack is robust and fit for real-world requirements. To further comprehensively evaluate our attack, we employed DP-SGD~\cite{abadi2016deep} as a defensive strategy to assess the attack's effectiveness. By reducing the model's ability to memorize training samples, DP-SGD defends against our attack. This finding is consistent with the outcomes observed in other attacks~\cite{carlini2023extracting,hu2023membership,duan2023diffusion}.

\paragraph{Contributions:} We make the following contributions.

\begin{itemize}[leftmargin=*]
    \item Many prior black-box attacks~\cite{hilprecht2019monte,chen2020gan,wu2022membership,zhang2024generated,matsumoto2023membership,dubinski2024towards} on image-generative models are no longer practical for the current generation of models and attack scenarios. We propose a black-box membership inference attack framework that is deployable against any generative model by leveraging the model's memorization of the training data.
    \item Consistent with the definition in Suya et al.~\cite{suya2023sok}, four attack scenarios are considered in which an attacker can perform an attack based on the \textit{query access} as well as the \textit{quality of the initial auxiliary data}, and three different attack models are used to determine the success rate of the attack, respectively. 
    \item The efficacy of the attack is evaluated on the CelebA, WIT, and MS COCO datasets using fine-tuned Stable Diffusion v1-5 as the representative target model. The attack's impact is analyzed by considering various factors: image encoder selection, distance metrics, fine-tuning steps, inference step count, member set size, shadow model selection, and the elimination of fine-tuning in the captioning model.
\end{itemize}
\paragraph{Roadmap.}
~\autoref{background} reviews key works on denoising generative models and membership inference attacks, including their application against diffusion models.~\autoref{Methodology} introduces our score-based black-box attack on diffusion models, tailored to four levels of attacker knowledge.~\autoref{experiment setup} describes our experimental setup, and~\autoref{Evaluation} compares our attack's effectiveness with existing methods and examines various influencing factors.~\autoref{defense} shows the effectiveness of our attacks against common defenses.~\autoref{related_work} discusses some other research related to our work.~\autoref{conclusion} concludes the paper, summarizing our main findings and contributions.

\section{Background} \label{background}

\subsection{Machine Learning}
In general, we can classify a machine learning model into discriminative (classification) models and generative models.

\subsubsection{Classification Models}
In the context of classification model training, the objective is to map an input $x$ to a category $y$. The functional representation of the model can be expressed as $y = \mathcal{M}(x)$, where $x$ denotes the input (e.g., an image), {$\mathcal{M}$ represents the classification model}, and $y$ denotes the corresponding label. The loss in the classification model, which quantifies the discrepancy between the predicted and true labels, can be articulated as follows:
\begin{equation*}
L(\theta) = \mathbb{E}_{x,y} \left[-\log(\mathcal{M}(x)_y)\right]
\end{equation*}
where $\theta$ denotes the parameters of $\mathcal{M}$,
$\mathcal{M}(x)$ denotes the model's output probability distribution over the possible categories, and $\mathcal{M}(x)_y$ specifically denotes the probability assigned to the correct label $y$.

\subsubsection{Generative Models} \label{bk:DGM} 

Generative models are designed to generate $\hat{x}= \mathcal{G}(z)$, where $z$ is the randomness not provided by users but inherent to the server hosting the {generator $\mathcal{G}$}. 

Popular generative models include VAEs~\cite{kingma2022autoencoding}, GANs~\cite{goodfellow2014generative}, and diffusion models~\cite{ho2020denoising}. 
Recently, diffusion models have gained significant traction. Building on the classical DDPM (Denoising Diffusion Probabilistic Models), a plethora of models, such as Imagen~\cite{saharia2022photorealistic}, DALL·E 3~\cite{BetkerImprovingIG}, GLIDE~\cite{nichol2022glide}, Stable Diffusion~\cite{rombach2022high}, have emerged and can generate high-quality images based on prompt information. In this paper, we mainly focus on diffusion models.

\subsection{Diffusion Models}
\subsubsection{Foundation of Diffusion Models} The diffusion model can be conceptualized as a process where a noisy image is incrementally denoised to eventually yield a high-resolution image. Given an image $x_0$,  the model initially imparts noise via $T$ forward (noisy-adding) processes. At timestep $t$, the noisy image $x_t$ can be represented as:
\begin{align}
    x_t = \sqrt{\bar{\alpha}_t} x_{0} + \sqrt{1 - \bar{\alpha}_t} \epsilon_t\label{eq:x_t} 
\end{align}
where $\bar{\alpha}_t = \prod_{i=1}^{t} \alpha_i$, and $\alpha_i$ is a predefined parameter that decreases incrementally within the interval $[0,1]$. The term $\epsilon_t$ is a random Gaussian noise derived using the reparameterization trick from multiple previous forward steps (more details in~\hyperref[appendix:diffusion]{Appendix~\ref*{appendix:diffusion}}).

The reverse process serves an objective opposite to that of the forward process. Starting from $\hat{x}_T=x_T$, upon obtaining the image $\hat{x}_t$ at timestep $t$, the reverse process aims to denoise it to retrieve the image $\hat{x}_{t-1}$. A neural network (e.g., U-Net) $\mathcal{U}_\theta$ is trained to predict the noise to be removed at each timestep. The loss function in the training process is defined as:
\begin{align}
    L_t(\theta) = \mathbb{E}_{x_0,\epsilon_t} \left[ \lVert \epsilon_t - \mathcal{U}_{\theta}(\sqrt{\bar{\alpha}_t} x_0 + \sqrt{1 - \bar{\alpha}_t} \epsilon_t, t) \rVert^2_2\right]
    \label{eq:diffusion-loss}
\end{align}

Alternatively, this loss function can also be employed to train DDIM~\cite{song2022denoising}, which has a deterministic reverse process.

\subsubsection{Prompt Guided Diffusion Models} 
Diffusion models~\cite{BetkerImprovingIG,ramesh2022hierarchical,saharia2022photorealistic,nichol2022glide,rombach2022high} mentioned above are also capable of generating high-quality images conditional on prompt information $p$, denoted as {$\hat{x} = \mathcal{G}(z,p)$} (further details can be found in~\hyperref[appendix: class-free]{Appendix~\ref*{appendix: class-free}}). Our experiments primarily utilize the current publicly available state-of-the-art model, Stable Diffusion~\cite{rombach2022high}. Distinct from other diffusion generative models~\cite{nichol2022glide,ramesh2022hierarchical,saharia2022photorealistic}, Stable Diffusion uniquely conducts both the forward and reverse processes within the latent space {(images simplified into lower-dimensional data)}. This approach offers advantages: the noise addition and removal processes operate over a smaller dimensionality, allowing for faster model training at lower computing costs. Additionally, within the latent space, the model can accommodate diverse prompt information to guide image generation. Importantly, Stable Diffusion is open-sourced and provides multiple high-quality pre-trained checkpoints online. This aligns well with the focus of our study on potential privacy concerns when fine-tuning pre-trained models for downstream tasks.



\subsection{Membership Inference Attacks} \label{mia}

Membership inference attacks (MIAs) primarily aim to determine whether a target data point $x$ is within the training dataset, often referred to as the \textit{member set}, of a given target model. The motivation behind these attacks is twofold: to ensure that models are not trained in a manner that misappropriates data and to safeguard against potential privacy breaches. MIA's underlying principle hinges on exploiting machine learning models' overfitting and memorization properties. Discerning the model's different reactions to member and non-member samples makes it feasible to infer the membership of the target point $x$. 

To formalize membership inference attacks, assume there is a data sample $x$, a model $\mathcal{M}_{\theta}$ trained with dataset $\mathcal{D}_m$. The attack $\mathcal{A}$ will access $\mathcal{M}_{\theta}$, $\mathcal{D}_m$ and take data sample $x$ as input. It will then output a bit $b\gets \mathcal{A}^{\mathcal{D}_m}(x, \mathcal{M}_\theta) \in\{0,1\}$ indicating whether $x$ was used in training (i.e., $x\in \mathcal{D}_m$) or not.
For simplicity, we use $\theta$ denoted model $\mathcal{M}_{\theta}$ and omit $\mathcal{D}_m$. 




Early MIAs predominantly target classification models and use the outputs from classifiers as the data to train their attack models~\cite{li2021membership,liu2022membership,salem2018mlleaks,shokri2017membership, hui2021practical,sablayrolles2019whitebox, long2018understanding, long2020pragmatic}.
Shokri et al.~\cite{shokri2017membership} introduced a technique for training shadow models designed to use shadow models to approximate the target model's behavior. {By collecting information from these shadow models, such as prediction vectors or training loss, as well as membership labels (e.g., members vs. non-members), adversaries can subsequently train a binary classifier. This classifier acts as an attack model to predict the membership of $x$ based on the data derived from querying $x$ on the target model.}

{Carlini et al.~\cite{carlini2022membership} argued that using loss as an attack feature is inadequate and constitutes a non-membership inference attack. Instead, the likelihood-ratio attack can serve as a better method. They first created two distributions, $\mathbb{D}_{\text{in}}$ and $\mathbb{D}_{\text{out}}$, based on the confidence scores of samples from the member and non-member sets, respectively.} Then, the distributions are used to calculate the probability density function of query data $x$ in the member set and non-member set. 

\begin{table}[t]
    \centering
    \caption{{The symbols \protect\whiteball[0.12cm] \halfball[0.12cm] and \blackball[0.12cm] represent an attacker's fully authorized, partially authorized, and unauthorized data access, respectively. Symbols \ding{51} and \ding{55} denote the use and non-use of a technique, respectively. `HP': stands for the model's parameter settings. `TD': training data used to train the target model. `IV': model's internal values, including loss and gradient. `IO': internal outputs (noisy images). `TSC': components (text and image) of the target sample. `SMs': whether the attack employs shadow models.}}
    \label{tab:Attack_model}
    \Large
    \resizebox{0.47\textwidth}{!}{
    \begin{tabular}{c|c|cccccc}
         \toprule[1pt] 
          & Method &  HP & TD & MIV & IOs & TSC & SMs \\ 
         \midrule   \multirow{4}{*}{\textbf{\begin{tabular}[c]{@{}c@{}}\rotatebox[origin=c]{90}{White} \end{tabular}}} &  Loss-based~\cite{hu2023membership} & \whiteball & \whiteball & \whiteball  & \whiteball  & \whiteball & \ding{55}\\ 
           & LiRA~\cite{carlini2023extracting} & \whiteball & \whiteball & \whiteball & \whiteball & \whiteball & \ding{51}  \\ 
             & LOGAN~\cite{matsumoto2023membership} & \whiteball & \whiteball & \whiteball  & \whiteball & \whiteball & \ding{55} \\ 
             & GSA~\cite{pang2023white} & \whiteball & \blackball & \whiteball & \whiteball & \whiteball & \ding{51}\\
             \hline \multirow{6}{*}{\textbf{\begin{tabular}[c]{@{}c@{}}\rotatebox[origin=c]{90}{Gray} \end{tabular}}} & SecMI~\cite{duan2023diffusion} & \blackball & \blackball & \blackball & \whiteball & \whiteball & \ding{51} \\
             & PIA~\cite{kong2023efficient} & \whiteball & \whiteball & \blackball & \whiteball & \whiteball & \ding{55}\\
             & PFAMI~\cite{fu2023probabilistic} & \whiteball & \blackball & \blackball & \whiteball & \whiteball & \ding{51}\\ 
             & {DRC}~\cite{fu2024model} & \whiteball & \whiteball & \blackball & \whiteball & \halfball
             & \ding{55}\\
             & {CLiD}~\cite{zhai2024membership} & \blackball & \blackball & \blackball & \whiteball & \halfball
             & \ding{51}\\
             & {Structure-Based}~\cite{li2024unveiling} & \whiteball & \whiteball & \blackball & \whiteball & \whiteball & \ding{55}\\
             \hline \multirow{8}{*}{\textbf{\begin{tabular}[c]{@{}c@{}}\rotatebox[origin=c]{90}{Black} \end{tabular}}} & GAN-Leaks~\cite{matsumoto2023membership} & \blackball &  \whiteball & \blackball & \blackball & \whiteball & \ding{55}\\ 
               & Intuition-attack~\cite{wu2022membership} & \blackball &  \whiteball & \blackball & \blackball & \whiteball & \ding{55}\\ 
               & {Pixel-attack}~\cite{dubinski2024towards} & \blackball &  \whiteball & \blackball & \blackball & \whiteball & \ding{55}\\ 
               & Distribution-attack~\cite{zhang2024generated} & \blackball &  \halfball & \blackball & \blackball & \whiteball & \ding{55}\\ 
               & Our Attack-I & \blackball &  \halfball & \blackball  & \blackball & \whiteball & \ding{51}\\ 
               & Our Attack-II & \blackball &  \halfball  & \blackball & \blackball & \halfball & \ding{51}\\
               & Our Attack-III & \blackball & \blackball & \blackball & \blackball & \whiteball & \ding{51}\\
               & Our Attack-IV & \blackball & \blackball & \blackball & \blackball & \halfball & \ding{51}\\
         \bottomrule[1pt] 
    \end{tabular}}
\end{table}

\paragraph{MIAs against Diffusion Models.} 
In the context of MIA against diffusion models, 
due to the structural differences between diffusion models and classification models, as well as the dissimilarities in their inputs and outputs, MIAs designed for classification models cannot be directly applied to diffusion models. 
The focus of the research lies in how to construct features for MIA.
We classify existing attacks against diffusion models as white-box, gray-box, and black-box, and introduce them separately. In white-box attacks, methods in this setting exploit the loss (derived from each timestep using~\autoref{eq:diffusion-loss}) and gradients (via backpropagation through the model). Gray-box attacks typically necessitate access to a model's intermediate outputs but do not require any internal model information. For gray-box attacks targeting diffusion models, the model's denoising trajectory, particularly the noisy images, is utilized as attack data. In contrast, black-box attacks operate without knowledge of the model's internal mechanics or process outputs, relying solely on the final generated images for analysis. In~\autoref{tab:Attack_model}, we compare all existing attacks. Each type of attack's details is deferred to~\autoref{black-box_attack} and~\autoref{related_work}.

\subsection{Problem Formulation}



\subsubsection{Threat Model} \label{threat model}
Given the query sample $x$ and black-box access to the target image-generative model $\mathcal{G}$, the goal of the attacker is to determine whether $x$ was used to train $\mathcal{G}$. More specifically, we focus on the {\it fine-tuning} process, namely, we care about the privacy of the fine-tuning dataset of $\mathcal{G}$, and do not care about the pre-training dataset.
We focus on fine-tuning because (1) the attacks will be similar for direct training, while the computational cost for experiments on fine-tuning MIA will be much smaller, and (2) the pretraining and fine-tuning paradigm is more popular with modern large models, and we will provide motivations for focusing on the fine-tuning process later.

We categorize the threat model into four scenarios (as shown in~\autoref{tab:Attack_model}) with two dimensions, namely:

\newcommand{\tuple}[1]{\langle #1 \rangle}
\begin{itemize}[leftmargin=*]

    \item \textbf{Target Sample Component.} 
     One distinct property of the current image generator models
     is that there exists the flexibility to input {text prompt} to guide model generation. 
    Two configurations for the attacker can be considered: First, the query data $x$ aligns with the training data as a text-image pair ($x = \tuple{T_q, I_q}$, where $T_q$ denotes the text component and $I_q$ denotes the corresponding image component). 
    Second, the attacker only obtains a suspect image potentially revealing private information without a corresponding caption ($x = \tuple{I_q}$). As our focus here is MIAs on text-to-image generative models, the scenario where $x$ solely consists of text is not deemed practical and hence, is not discussed.
    
    \item \textbf{Auxiliary Dataset.} 
    Similar to all other MIAs, we assume an auxiliary dataset $\mathcal{D'}$ is available. It is used to train the shadow models $\mathcal{G}^s$ to mimic the behavior of the target model. 
    {We consider two scenarios for $\mathcal{D'}$, indicating whether the auxiliary dataset overlaps with the training set of the target model. In the case of the first scenario, $\mathcal{D'}$ contains $50\%$ real samples that were used to fine-tune the target model. In contrast, the second scenario represents $\mathcal{D'}$ is sampled from the same distribution as the target model's training set but without any overlap.}
\end{itemize}

\subsubsection{Motivation}

{Our work aims to reveal privacy infringements in the datasets used for fine-tuning image generative models.} Given the full open-source nature of the Stable Diffusion~\cite{rombach2022high}, and the extensive availability of pre-trained models capable of generating photorealistic images from entities like CompVis\footnote{\url{https://huggingface.co/CompVis/stable-diffusion}} and Stability AI\footnote{\url{https://github.com/Stability-AI/generative-models}}, there has been an increasing trend of leveraging these pre-trained models for fine-tuning to specific downstream tasks. Furthermore, an increasing number of companies, such as Amazon\footnote{\url{https://aws.amazon.com/sagemaker/jumpstart/}}, OctoML\footnote{\url{https://octoml.ai/blog/the-beginners-guide-to-fine-tuning-stable-diffusion/}}, and CoreWeave\footnote{\url{https://docs.coreweave.com/cloud-tools/argo}}, are offering services in this domain. The data privacy issue during the training of these downstream tasks has not been explicitly studied. Our work seeks to uncover privacy violations in this process and raise awareness of them.



\begin{table}[t]
        \centering
        \caption{Results of utilizing our attack as an auditing tool. `Member' refers to the similarity between generated images mimicking a specific artist or style and the actual works of that artist or style. The `Non-member' sample similarity score is calculated by querying the model with samples not seen during fine-tuning and comparing them to their ground-truth images. `Diff.' refers to the difference in similarity scores between member samples and non-member samples. Additionally, the attack results demonstrate the effectiveness of the attack using a shadow model trained on an auxiliary dataset. The similarity scores are calculated as an average of $3$ generated images. The art style with the highest attack accuracy is highlighted in bold.}
        \label{tab:case_study}
        \centering
        \resizebox{0.48\textwidth}{!}{
        \centering
        \begin{tabular}{ccccc}
        \toprule
         Art Style (Artist) & Member & Non-member & Diff. & ROC-AUC\\ \midrule
         Vincent van Gogh & $\bm{0.92}$ &  $\bm{0.44}$ & $\bm{0.48}$ & $\bm{0.91}$\\
         Baishi Qi& $0.77$ &  $0.40$ & $0.37$ & $0.88$\\
         Ukiyo-e & $0.70$ & $0.45$ & $0.25$ & $0.87$\\
         Uemura Shoen & $0.88$ & $0.41$ & $0.47$ & $0.89$ \\
         Wanostyle & $0.80$ & $0.49$ & $0.31$ & $0.87$\\
         Ken Kelly & $0.80$ & $0.47$ & $0.33$ & $0.89$\\
         Shanshui Painting & $0.88$ & $0.47$ & $0.41$ & $0.86$ \\
        \bottomrule
        \end{tabular}}
\end{table}

\subsubsection{Case Study}

{To better demonstrate the risks of data misuse during the fine-tuning process and the ability of models to steal artistic styles, we designed a simple case study in this section. In the study, we collected seven models from Civitai\footnote{\url{https://civitai.com/}}, a website that shares fine-tuned models. These models were utilized to generate images that mimic the styles of specific artists. We define synthesized samples that mimic the same art style as the fine-tuning data as `Member' and those unrelated to the fine-tuning set as `Non-member'. The `Member' similarity score is calculated by comparing the generated images with artworks by the same artists. In contrast, the `Non-member' similarity score is computed by querying the model with data not used during fine-tuning and measuring the similarity between the generated and the ground-truth images. We present the `Member' and `Non-member' similarity scores for each of the seven models in~\autoref{tab:case_study}. In the case study, Cosine similarity served as the distance metric, and \texttt{DeiT} was employed to extract the features. For each model, we collected $60$ samples for both member and non-member categories. Each sample was queried three times, and the average of these three similarity scores was used as the final similarity score for that sample.}

{According to~\autoref{tab:case_study}, the similarity score of `Member' is at least $0.25$ higher than `Non-member' samples. In some cases, such as when imitating Vincent van Gogh's works, the difference in similarity between `Member' and `Non-member' samples can reach up to $0.48$. The distinct difference in similarity scores demonstrates that the model can copy the relevant artistic style after being trained on an artist's work.}

\subsubsection{Auditing Tool}

{From the perspective of artists, it is concerning that models can steal their artistic styles after being trained on their works. To protect their copyright, artists need an auditing tool to detect suspicious models that may have used their work without authorization. Such an auditing tool is crucial today, as websites like Civitai already contain many fine-tuned models capable of imitating artistic styles and anime characters. Our work is based on the model's memorization of samples during training. By incorporating the objective function of diffusion models (more details can be found in~\autoref{Methodology}), we designed a score-based membership inference attack. As observed in~\autoref{tab:case_study}, the significant difference between the `Member' and the `Non-member' similarity score suggests that our attack can serve as an auditing tool to detect potential misuse of training data.}

{To evaluate the feasibility of our attack as an auditing tool and ensure consistency with subsequent experiments, we used the shadow model technique to attack each target model (model from Civitai). We set the the auxiliary dataset does not overlap with the member and non-member sets of the target model. Additionally, the sizes of all member and non-member sets are identical. Then, we trained an MLP as the attack model using the shadow model's member and non-member data, and we present the attack ROC-AUC in~\autoref{tab:case_study}. The preliminary results of these attacks undoubtedly demonstrate the feasibility of our approach.}



\subsection{Existing Solutions}
\label{subsec:existing}

\subsubsection{Black-box MIA against Traditional Image-generatve Models} \label{black-box_attack}
There are existing black-box MIAs targeting VAEs and GANs. They share a similar underlying idea, which is that if the target sample $x$ was used during training, the generated samples would be close to $x$. Monte-Carlo attack~\cite{hilprecht2019monte} invokes the target model many times to generate many samples first. Given $x$, it measures the number of generated samples within a specific radius. The more samples there are, the higher the likelihood that $x$ is part of the member set.  

GAN-Leaks~\cite{chen2020gan} employs a similar intuition, using the shortest distance of the generated samples from the target sample as the criterion. It also proposes another attack assuming an extra ability to optimize the noise input $z$ to the generator (which is not strictly the black-box setting; we will describe it and compare with it in the evaluation) so it can reduce the number of generated samples. More formal details about these attacks are deferred to~\hyperref[appendix: discussion]{Appendix~\ref*{appendix: discussion}}.

The reason we cannot apply Monte-Carlo attack~\cite{hilprecht2019monte} and GAN-Leaks~\cite{chen2020gan} to diffusion models is that both attack methods require the model to sample a large number of images. Diffusion models progressively denoise during the inference process, involving dozens of steps, unlike VAEs and GANs, which require only a single step. Both Monte-Carlo attack and GAN-Leaks need to construct $100K$ samples to achieve optimal attack performance~\cite{chen2020gan}. For the diffusion model, this will take even hundreds of times longer in terms of computing time. Furthermore, these attacks are unsuitable for conditional generative models. Although GAN-Leaks proposed a partial-black attack, conditional embedding (e.g., text embedding) in diffusion models is significantly more complex than the initial noise $z$ in GANs and VAEs. Therefore, traditional black-box MIAs are not feasible for current diffusion models.

\subsubsection{Black-box MIA against Recent Diffusion Models} \label{black-box_attack_recent}
Matsumoto et al.~\cite{matsumoto2023membership} directly adopted the concept of GAN-Leaks~\cite{chen2020gan} to diffusion models. However, as diffusion models are more complex, the attack is bottlenecked by the time required to sample a large number of samples. 

Wu et al.~\cite{wu2022membership} leveraged the intuition that the generated samples exhibit a higher degree of fidelity in replicating the training samples, and demonstrate greater alignment with their accompanying textual description. However, the authors {did not use the shadow model technique and} only tested their attack on off-the-shelf models with explicitly known training sets. In the realistic setting where the training set is unknown (which is the purpose of MIAs), their attack cannot work.

{Dubinski et al.~\cite{dubinski2024towards} designed their attack against API-based generative machine
learning services (e.g., Midjourney\footnote{\url{https://www.midjourney.com/home}}) by directly comparing the pixel-level error between generated samples and known training samples. However, similar to Wu et al.~\cite{wu2022membership}, they did not use the shadow model technique in the black-box scenario (they trained shadow models in white-box settings) and assumed the attacker already knew the training dataset (LAION Aesthetics v2.5+), using it as the member set. This attack assumes the attacker can access excessive information, making it impractical in real-world scenarios.}

Additionally, Zhang et al.~\cite{zhang2024generated} trained a classifier based on samples generated by the target model (labeled $1$) and samples not used in training (labeled $0$). The classifier can then determine whether the target sample was used in training. However, it needs to (1) know the non-training samples, and (2) ensure the two distributions (of generated samples and non-training samples) are different enough. Both conditions are not necessarily true in a realistic setting.

\section{Methodology}\label{Methodology}

{In this section, we introduce our attack, which is based on the model's memorization of training samples. Current black-box membership inference attacks on generative models, such as GAN-Leaks~\cite{chen2020gan} and Monte-Carlo attacks~\cite{hilprecht2019monte}, also exploit this characteristic.} However, GAN-Leaks relies on Parzen window density estimation to estimate the probability of query samples~\cite{parzen1962estimation} that belong to the training set. This method often results in unstable probability estimates due to the large sampling size, as we mentioned in~\autoref{black-box_attack}. We propose utilizing the {\it intrinsic characteristics of diffusion models with formal proofs} to design a more efficient and suitable attack for diffusion models. {Specifically, we leverage the training objective of diffusion models to more directly and intuitively quantify the model's memorization of query samples using similarity scores. Based on the results of the similarity score analysis, we determine the membership of query samples.}

\subsection{Theoretical Foundation} \label{theoretical_foundation}

We aim to establish a detailed theory demonstrating the similarity score between the query image $I_q$ and generated image $I_g$ can be used as a metric to infer the membership of $x$. {It is important to note that a high similarity score between $I_q$ and $I_g$ indicates a low distance between the two images.} We leverage the internal property of the diffusion model, which is inherently structured to optimize the log-likelihood: If $x$ is in the training set, its likelihood of being generated should be higher. However, due to the intractability of calculating log-likelihood in diffusion models, these models are designed to use the Evidence Lower Bound (ELBO) as an approximation of log-likelihood~\cite{ho2020denoising}, as shown later in~\autoref{eq:likelihood}. 
In~\autoref{theorem_1}, we first argue that ELBO of the diffusion model can be interpreted as a chain of generating images at any given timestep that approximates samples in the training set. Then, in~\autoref{theorem_2}, based on the loss function of the Stable Diffusion~\cite{rombach2022high}, we extend the result and demonstrate that this argument remains valid. Therefore, we can reasonably employ the similarity between the generated images and the query image as our attack.
Note that GAN-Leaks~\cite{chen2020gan} also shares this intuition of using similarity. However, it relies more on intuition and lacks a solid foundation, as the training of GANs is different (not a streamlined process as in diffusion).

From the perspective of the training process, we proposed these two theorems that facilitate our attack.

\begin{theorem} \label{theorem_1}
    Assuming we have a pre-trained diffusion model $\hat{x}_{\theta}$\footnote{We previously use $\mathcal{U}_{\theta}$ to denote U-Net, now by slightly abusing notations we use $\hat{x}_{\theta}$ for easier presentations.} with its training set $\mathcal{D}_m$, and use a bit $b$ to represent the membership of query sample $x$ {($1$ for member and $0$ for non-member)}. 
    The higher similarity scores between the query data $x$ and its generated image $\hat{x}_{\theta}(x_t, t)$, the higher the probability of $\Pr{b=1 | x, \theta}$. 
    \begin{equation*}
        \Pr{b=1 | x, \theta} \propto -{\left\| x_0 - \hat{x}_{\theta}(x_t,t)\right\|^2_2}
    \end{equation*}
    where $\theta$ denotes the parameters of the model.
\end{theorem}
 
{\begin{proof}[{Proof Sketch}]
We first demonstrate that diffusion models use ELBO to approximate the log-likelihood of the training dataset. By restructuring the optimization function, we find that the diffusion model primarily focuses on predicting the noise $\epsilon_t$ at $t$-th step. Using~\autoref{eq:x_t}, we show that the objective function of the diffusion model can also be expressed in terms of predicting $\hat{x}$ at each step. Therefore, a data sample from the diffusion model's member set is expected to have higher similarity with its replication $\hat{x}_{\theta}(x_t,t)$ at each step. As a result, the denoised sample from the diffusion model should naturally exhibit higher similarity scores with member set samples. The full proof can be found at~\hyperref[appendix:theorem1_proof]{Appendix~\ref*{appendix:theorem1_proof}}. 
\end{proof}}

In the above, we have linked the probability of query sample $x$ belonging to the member set to its similarity score with generated images in the unconditional diffusion model. For this type of diffusion model, although we can prove the training image has this property with its replica. We still cannot design the black-box attack on it because the inference process is random. We cannot control the unconditional diffusion model to reconstruct the specific data sample. This generation process is the same with VAEs and GANs. Hence, the existing black box attacks are to sample a large number of images from the models~\cite{matsumoto2023membership}. And then do the Monte Carlo~\cite{hilprecht2019monte} or GAN-Leaks~\cite{chen2020gan} attack.

However, we can employ this property to execute the membership inference attack with conditional diffusion models (e.g., Stable Diffusion). {The main difference between conditional and unconditional diffusion models is that the former can perform conditional generation.} According to the prompt input, Stable Diffusion can generate an image that aligns with it. Therefore, we can use prompts to guide the model and synthesize images for a specific data sample. {In~\autoref{theorem_2}, we prove this property valid in the Stable Diffusion.}


\begin{theorem} \label{theorem_2}

For a well-trained Stable Diffusion model{\footnote{In our work, we used the pre-trained Stable Diffusion-v1-5 from CompVis, which was trained for $150,000$ A100 hours.}}, $\hat{z}_{\theta}$\footnote{$\hat{z}_{\theta}$ represents only the U-Net in Stable Diffusion, excluding the VAE and text encoder.}, the query sample is $x$, and the membership of $x$ is denoted as $b$ ($1/0$ for member/non-member). {$\mathrm{D}/\mathrm{E}$ refers to the decoder/encoder module of the VAE in Stable Diffusion. A pre-trained text encoder, $\phi_{\theta}$, converts the input conditional prompt $p$ into the text embedding that guides image generation. The similarity scores remain a viable metric for assessing the membership of query data $x$.} This relationship can be expressed in the following mathematical formulation:
\begin{equation*}
        \Pr{ b=1 | x, \theta} \propto -{\left\| \mathrm{D}(z_0) - \mathrm{D}(\hat{z}_{\theta}(z_t, t, \phi_{\theta}(p)))\right\|^2_2} 
\end{equation*}
Where $z_t$ represents the latent representation, $z_0 = \mathrm{E}(x)$. 
\end{theorem}

{\begin{proof}[{Proof Sketch}]To establish~\autoref{theorem_2}, we begin by examining the loss function of Stable Diffusion. We find that the optimization objective and the diffusion process in Stable Diffusion remain consistent with the unconditional diffusion model. However, the diffusion/denoising process is moving from the pixel level to the latent space. Through reinterpreting the noise prediction $\epsilon_t$ at each step, the optimization objective of Stable Diffusion can also be viewed as predicting the initial latent variable $z_0$ at each step. By incorporating the Decoder $\mathrm{D}$, we prove that in Stable Diffusion, the member sample $\mathrm{D}(z_0)$ should have a higher similarity score with its replicate $\mathrm{D}(\hat{z}_{\theta}(z_t, t, \phi_{\theta}(p)))$. The detailed proof of~\autoref{theorem_2} is presented at~\hyperref[appendix:theorem2_proof]{Appendix~\ref*{appendix:theorem2_proof}}.
\end{proof}
}

Considering the realistic situation and settings, we designed four attacks (as shown in~\autoref{sec:bs_attack}) to use this property in different scenarios. However, for general representation, we simplify denote the image generated by the model as $I_g$ (which also corresponds to $\hat{x}$ in~\autoref{bk:DGM}, {$\hat{x}_{\theta}(x_t,t)$ in~\autoref{theorem_1}, and $\mathrm{D}(\hat{z}_{\theta}(z_t, t, \phi_{\theta}(p)))$ in~\autoref{theorem_2}}). The similarity score between $I_g$ and $I_q$ from the query data $x$ can be represented as $S(I_q, I_g)$. Here, $S$ is a distance metric (e.g., Cosine similarity, $\ell_1$ or $\ell_2$ distance, or Hamming distance). Given that a higher similarity score (low distance) indicates a higher probability of the data being a training sample, the inference model can be formulated accordingly.
\begin{equation}
    \mathcal{A}_{base}(x,\theta) = \mathbbm{1} \left\{ S(I_q, I_g) \geq \tau \right\}
\end{equation}

The base inference model relies on computing the similarity scores between $I_g$ and $I_q$. {If the similarity score $S(I_q, I_g)$ exceeds a certain threshold, the inference model will determine that the data record $x$ associated with $I_q$ comes from the member set.}

\subsection{Attack Pipeline}

{According to~\autoref{theoretical_foundation}, our attack needs to calculate the similarity between query image $I_q$ and generated image $I_g$. We choose to compute the image embedding similarity scores by using image feature extractors. Also, to execute our attack on query data that lacks text components, we incorporate the captioning model in our work. Our work seeks to uncover privacy violations in this process and raise awareness of them.}  

\paragraph{Image Feature Extractor.} As we follow the high-level intuition of GAN-Leaks and use image similarities to determine membership, { we employ distance metrics (e.g., Cosine similarity, $\ell_1$ or $\ell_2$ distance, or Hamming distance) to formally quantify this similarity.} It has been observed that the semantic-level similarities are substantially more effective than pixel-level similarities~\cite{wu2022membership}. Therefore, we utilize a pre-trained image encoder (i.e., \texttt{DETR}, \texttt{BEiT}, \texttt{EfficientFormer}, \texttt{ViT}, \texttt{DeiT}) to extract semantic representations from the images. 

\paragraph{Captioning Model.} In our work, under certain scenarios, the query data $x$ may lack the text component $T_q$ and only include $I_q$. Consequently, we resort to a captioning model to generate the corresponding text. For our experiments, we utilize \texttt{BLIP2}~\cite{li2023blip} as the captioning model. To ensure that the generated textual descriptions closely match the style of the model's training dataset, we also consider further use of the auxiliary dataset to fine-tune the captioning model. 


\begin{algorithm}[t]
    \caption{High-level Overview of Our Attack.}
    \label{alg:High-level}
    \begin{algorithmic}[1]
    \Require Query sample $x$, target model $\mathcal{G}$, distance metrics $S(\cdot,\cdot)$, the image captioning model $\mathcal{C}$, the instantiation of attack $\mathcal{A}$, the statistical function $f$, and the image feature extractor $E$.
    \If{$T_q$ $\notin$ $x$} {\Comment{Check for text components in $x$.}}
    \State $T_q$ = $\mathcal{C}(I_q)$ \Comment{Synthesize the text for $\mathcal{G}$.}
    \EndIf
    \For{$i=1$ \textbf{to} $m$} \Comment{Perform $m$ repetitive queries.}
        \State $I_g^{i}$ = $\mathcal{G}(T_q)$ 
    \EndFor
    \Ensure $\mathcal{A}( f \left[\tuple{S(E(I_q),E(I_g^{i}))}_{i = 1}^{m}\right])$\Comment{MIA results.}
    \end{algorithmic}
\end{algorithm}

\paragraph{Attack Overview.}~\autoref{alg:High-level} gives the high-level overview of our attack. The intuition is to compare the generated images with the query image and compute a similarity score used for MIAs (specific instantiations of $\mathcal{A}$ to be presented in \autoref{sec:instantiation}).
Depending on whether the text is available or not, we might need the captioning model to synthesize the text. {Once the captioning is complete, we repeatedly query the target model $m$ times for each query image, then apply a statistical function $f$ (e.g., mean, median) to aggregate the $m$ similarity score vectors for each query image.} Finally, we return the aggregated similarity scores to determine the target/query data's membership.  

Note that while the attack pipeline is perhaps straightforward, its intuition relies on the formal analysis of the diffusion models. We first describe its theoretical foundation and then instantiate it with different MIA paradigms based on the output score in the following. 

\subsection{Instantiations}
\label{sec:instantiation}
Utilizing the scores obtained from~\autoref{alg:High-level}, we instantiate three different types of MIAs according to \autoref{mia}. In our evaluation, we try all three of them, and observe the last one is usually the most effective one.

\paragraph{Threshold-based Membership Inference Attack.} {Since the threshold-based MIA uses a scalar for comparison, the similarity scores obtained after applying $f$ are calculated for each image patch (e.g., \text{ViT} generate $196$ patches, more details in~\hyperref[appendix:framework]{Appendix~\ref*{appendix:framework}}). Therefore, to compute the overall image similarity, these similarity scores need to be averaged, i.e.,}
\begin{align}
    \displaystyle \frac{1}{k} \sum_{j=1}^{k} f \left[\left\langle S \left( E(I_q), E(I_g^i) \right) \right\rangle_{i = 1}^{m}\right]_{j}\geq \tau\label{equ:our_attack} 
\end{align}
{Where $k$ refers to the patch size used by the image feature extractors, $S$ represents the distance metrics, and $E$ denotes the image feature extractor. It is important to note that $\tau$ is determined in advance using member and non-member samples from the shadow model. Specifically, when the statistical function $f$ is mean, we calculate each query sample's average feature similarity score, then average these scores across all patches and scale them (using Min-Max scaling~\cite{patro2015normalization}) to the range $[0,1]$. After scaling, we use Youden's index~\cite{Youden1950} to determine the best threshold $\tau$ that yields the highest AUC. This $\tau$ is then used to attack the target model.}



\paragraph{Distribution-based Membership Inference Attack.}
Following the work by Carlini et al.~\cite{carlini2022membership}, we know we can also use the likelihood ratio attack against diffusion models. In our analysis, we leverage similarity scores derived from shadow models to delineate two distinct distributions: $\mathbb{Q}_{in}$ and $\mathbb{Q}_{out}$.
Specifically:
    For $\mathbb{Q}_{in}$, consider image $I$ that belong to the member set $\mathcal{D}_m$. We then define $\mathbb{Q}_{in}$ as 
    \begin{align*}
    \mathbb{Q}_{in} &= \left\{ f \left[\left\langle S \left( E(I), E(I_g^i) \right) \right\rangle_{i = 1}^{m} \right]\;\bigg|\; I \in \mathcal{D}_m \right\}.
    \end{align*}
    
    Similarly, for $\mathbb{Q}_{out}$, when image $I$ are part of the non-member set $\mathcal{D}_{nm}$, we have 
    \begin{align*}
    \mathbb{Q}_{out} &= \left\{ f \left[\left\langle S \left( E(I), E(I_g^i) \right) \right\rangle_{i = 1}^{m}\right] \;\bigg|\; I \in \mathcal{D}_{nm} \right\}.
    \end{align*}

For target query point $I_q$, membership inference can be deduced by assessing:
\[
\mathsf{Pr}\left[ f\left[ \left\langle S \left( E(I_q), E(I_g^i) \right) \right\rangle_{i = 1}^{m}\right] \bigg| \mathbb{Q}_{in} \right]
\]
and
\[
\mathsf{Pr}\left[ f \left[\left\langle S \left( E(I_q), E(I_g^i) \right) \right\rangle_{i = 1}^{m}\right] \bigg| \mathbb{Q}_{out} \right]
\]

\paragraph{Classifier-based Membership Inference Attack.} Given that the obtained similarity score is represented as a high dimensional vector, 
the classifier-based MIA feeds
$ f \left[\left\langle S \left( E(I_q), E(I_g^{i}) \right) \right\rangle_{i = 1}^{m}\right]$ 
directly into a classifier (we use a multilayer perceptron in our evaluation). This approach aligns with the methods of Shokri et al.~\cite{shokri2017membership}, leveraging the machine learning model as the inference model to execute the attack.

In evaluation, although we can use different functions of $f$, we observe a simple $f$ that takes the mean of all $m$ similarity scores performs pretty stable, so we just use the mean function for all three MIAs throughout the evaluation.

\section{Experiment Setup} \label{experiment setup}
\subsection{Datasets} \label{experiment dataset}
Stable Diffusion v1-5 is pre-trained on LAION-2B~\cite{schuhmann2022laion5b} and LAION-Aesthetics. To guarantee the integrity and effectiveness of our work, we utilize the MS COCO~\cite{lin2015microsoft}, CelebA-Dialog~\cite{jiang2021talkedit}, and WIT datasets~\cite{srinivasan2021wit} for evaluation, ensuring that there is no overlap with the pre-training dataset. {We label the samples in the member set as the positive class and the non-member samples as the negative class.}


\paragraph{MS COCO}is a large-scale dataset featuring a diverse array of images, each accompanied by five similar captions, amounting to a total of over $330$k images. The MS COCO dataset~\cite{lin2015microsoft} has been extensively utilized in various image generation models, including experiments on DALL·E 2~\cite{ramesh2022hierarchical}, Imagen~\cite{saharia2022photorealistic}, GLIDE~\cite{nichol2022glide}, and VQ-Diffusion~\cite{gu2022vector}. In this work, we randomly selected $50$k images along with their corresponding captions to do the experiments. Each image is paired with a single caption to fine-tune the model.

\paragraph{CelebA-Dialog}is an extensive visual-language collection of facial data. Each facial image is meticulously annotated and encompasses over $10,000$ distinct entities. Given that each face image is associated with multiple labels and a detailed caption, the dataset is suitable for a range of tasks, including text-based facial generation, manipulation, and face image captioning. Facial information has consistently been regarded as private; hence, utilizing CelebA-Dialog~\cite{jiang2021talkedit} in this study aligns with our objective of detecting malicious users fine-tuning the Stable Diffusion model~\cite{rombach2022high} for simulating genuine face generation.

\paragraph{WIT}is a vast image-text dataset encompassing a diverse range of languages and styles of images and textual descriptions. It boasts $37.6$ million image-text pairs and $11.5$ million images, showcasing remarkable diversity. We leverage this dataset specifically to evaluate the robustness of our attack in handling such heterogeneous data.


\begin{table}[t]
    \centering
    \caption{The default parameters used in~\autoref{Evaluation}.}
    \label{tab:default_setting}
    \resizebox{0.45\textwidth}{!}{
    \begin{tabular}{cc}
    \toprule[0.9pt]
        Parameters & Experiment setting for our work \\
        \midrule[0.5pt]
        Training data size & $100$  \\
        Epoch number & $500$ \\
        Resolution & $512\times512$  \\
        Batch size & $4$ \\
        Learning rate & $5\times10^{-5}$\\
        Gradient accumulation steps & $4$ \\
        Inference step &  $30$ \\
        Image feature extractor & \texttt{DeiT} \\
        Captioning model & \texttt{BLIP2}\\
        Distance metrics & Cosine similarity \\
        Attack type & Classifier-based\\
    \bottomrule[0.9pt]
    \end{tabular}}
\end{table}

\subsection{Evaluation Metrics}

To systematically evaluate the efficacy of our proposed attack, we opted for multiple evaluation metrics as performance indicators. {Similar to other comparable attacks~\cite{duan2023diffusion,hu2023membership,matsumoto2023membership,kong2023efficient,wu2022membership,carlini2023extracting,carlini2022membership}, we employ ASR (Accuracy of Membership Inference), Area Under the ROC Curve (AUC), and True Positive Rate (TPR) at low False Positive Rate (FPR) as our evaluation metrics.} In~\autoref{Evaluation}, all experiments are evaluated under the condition that the member set and non-member set have the same size.

We opted to use Stable Diffusion v$1$-$5$\footnote{\url{https://huggingface.co/runwayml/stable-diffusion-v1-5}} checkpoints as our pre-trained models. {The fine-tuning code script was modified from the Huggingface Diffusers package\footnote{\url{https://huggingface.co/docs/diffusers/v0.9.0/en/training/text2image}}.} All experiments were carried out using two Nvidia A100 GPUs, and each fine-tuning of the model required an average of three days. We presented the default fine-tuning and attack settings in~\autoref{tab:default_setting}.


\subsection{Baseline Attacks} \label{sec:bs_attack}
For our evaluation, we first compare our work with existing black-box attacks on diffusion models~\cite{matsumoto2023membership,fu2023probabilistic}. 
For our attack, based on the categorization provided in~\autoref{threat model}, the attacker will obtain information of two distinct dimensions, leading to four different scenarios. We call them Attack-I to Attack-IV. Below we introduce them in more detail.

\paragraph{Matsumoto et al.~\cite{matsumoto2023membership}} employed the full-black attack framework from GAN-Leaks.

\paragraph{Zhang et al.~\cite{zhang2024generated}} utilized a novel attack strategy involving a pre-trained ResNet18 as a feature extractor. This approach focuses on discriminating between the target model's generated image distribution and a hold-out dataset, thereby fine-tuning ResNet18 to become a binary classification model.

\myparagraph{$(x=\tuple{T_q,I_q}, \mathcal{D'}\cap\mathcal{D}_m \neq \emptyset)$ }
In this attack scenario, we assume the attacker has access to partial samples from the actual training (fine-tuning) set of the target model (attacker's auxiliary data $\mathcal{D'}$ overlaps with the fine-tuning data $\mathcal{D}_m$). Furthermore, $x$ includes both the image and the corresponding text (caption information). An attacker can directly utilize $T_q$ to obtain $I_g$, then employ the similarity between $I_g$ and $I_q$ to ascertain the membership of $x$.




\myparagraph{$(x=\tuple{I_q},\mathcal{D'}\cap\mathcal{D}_m \neq \emptyset)$}
In this scenario, 
the attacker does not possess a conditional prompt that can be directly fed into the target model. The attacker needs to use an image captioning model to produce a caption for $I_q$. This caption is subsequently used as the input for $\mathcal{G}$. The process culminates in the computation of similarity between the query image $I_q$ and the image generated by $\mathcal{G}$.


\myparagraph{$(x=\tuple{T_q,I_q}, \mathcal{D'}\cap\mathcal{D}_m = \emptyset)$} is similar to the first scenario (the difference is the attacker's auxiliary dataset does not intersect with the target training dataset). The attack (as shown in \autoref{alg:High-level}) is the same, but we expect a lower effectiveness.

\myparagraph{$(x=\tuple{I_q},\mathcal{D'}\cap\mathcal{D}_m = \emptyset)$} is similar to the third scenario {(there is no overlap between the attacker's auxiliary dataset and the target member set). This attack represents the hardest situation, and we think it will get the lowest accuracy.}

\begin{figure*}
    \centering
    \includegraphics[width=0.90\textwidth]{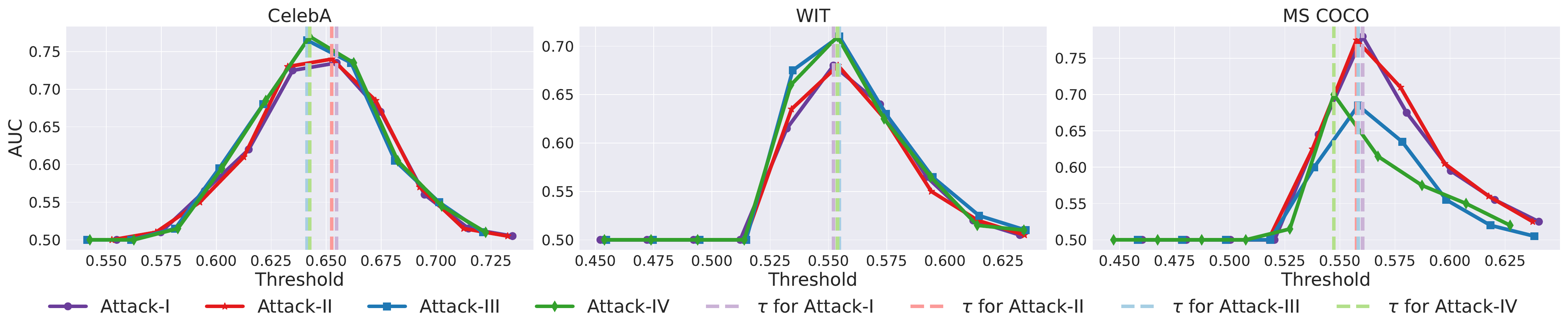}
    \caption{{Impact of different threshold values on attack results using the ROC-AUC metric across three datasets. Here, $\tau$ represents the best threshold selected for each attack from the shadow model based on the AUC scores.}}
    \label{fig:diff_threshold}
\end{figure*}

\section{Experiments Evaluation} \label{Evaluation}

\subsection{Comparison with Baselines}\label{sec:baseline}
\begin{table}[t]
    \vspace{0.2cm}
    \caption{Comparison between the attacks by Zhang et al.~\cite{zhang2024generated}, Matsumoto et al.~\cite{matsumoto2023membership} (applying GAN-Leaks against the diffusion model) versus our methods. The best attack result is highlighted in bold.}
    \label{tab:baseline}
    \centering
    \resizebox{0.40\textwidth}{!}{
    \begin{tabular}{cccc}
    \toprule[0.5pt]
      \multirow{2}{*}{Attack type}   &  \multicolumn{3}{c}{CelebA-Dialog} \\ \cmidrule[0.3pt](lr){2-4}
         & ASR  & AUC & TPR@FPR=1\% \\
         \midrule[0.3pt]
         Matsumoto et al.~\cite{matsumoto2023membership} & $0.52$ & $0.50$ & $0.01$\\
         Zhang et al.~\cite{zhang2024generated} & $0.51$ & $0.49$ & $0.01$ \\
         \hdashline[1pt/2pt]
         Attack-I & $0.85$ & $0.93$ & $0.53$ \\
         Attack-II  & $\bm{0.88}$ & $\bm{0.93}$ & $\bm{0.60}$ \\
         Attack-III & $0.87$ & $0.94$ & $0.54$ \\
         Attack-IV & $0.87$ & $0.93$ & $0.57$ \\
    \bottomrule[0.5pt]
    \end{tabular}}
\end{table}

\begin{figure*}[t!]
    \centering
    \includegraphics[width=0.95\textwidth]{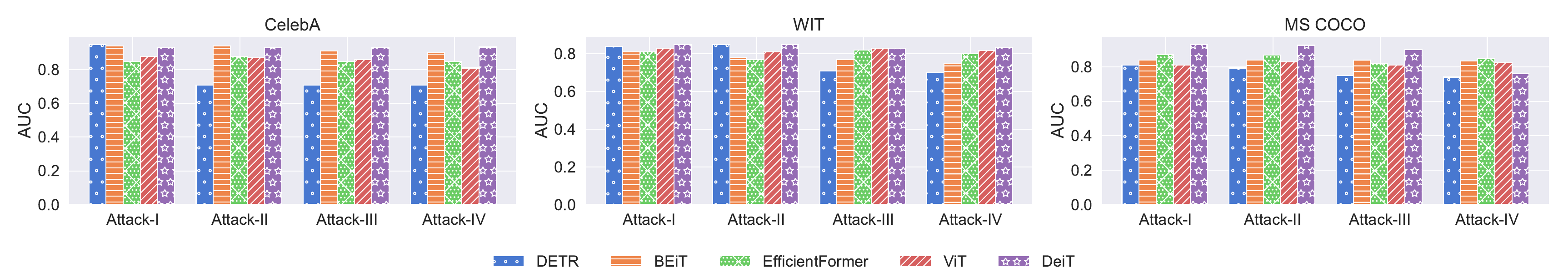}
    \caption{AUC results on three datasets and four attack scenarios comparing five different image feature extractors.}
    \label{fg:encoder}
\end{figure*}

Results are shown in~\autoref{tab:baseline}. We ensure consistency in simulating real-world scenarios, wherein the number of images that a malicious publisher can sample from the target generator is limited. Under the constraint of limited sample size, we observe that the accuracy of both baseline attacks nearly equates to random guessing. We conjecture that this is due to their reliance on a large number of synthesis images for decision-making. Specifically, Zhang et al.~\cite{zhang2024generated} requires learning the distributional differences between generated image samples and non-member samples using ResNet18, based on a substantial volume of images sampled from the target model, and subsequently applying this knowledge to assess the input query data. However, such an attack premise falters in realistic scenarios where a malicious model publisher restricts the number of images a user can obtain from the model, preventing attackers from sampling a large volume of images to conduct the attack. Under such constraints, the effectiveness of attacks by Zhang et al.~\cite{zhang2024generated} and others is inevitably compromised, as the insufficient sample size hampers the ability to accurately discern the differences between the two data distributions. Similarly, the approach by Matsumoto et al.~\cite{matsumoto2023membership} encounters a hurdle; in scenarios of limited generative sample availability, it becomes challenging to find a suitable reconstruction counterpart and calculate its distance from the original data record. Consequently, these methods fail to achieve high attack success rates under sample-restricted conditions. In contrast, the four attacks we propose still attain a high success rate despite the limited number of generative samples. This is attributed to our attacks being based on the similarity scores as proposed in~\autoref{theoretical_foundation}, which, while influenced by the quality of the model's generated images, is not hindered by the quantity of these images.

\subsection{Impact of Different Thresholds}

{In our work, we introduced three different types of attacks: \textit{threshold-based}, \textit{distribution-based}, and \textit{classifier-based} attacks. The \textit{threshold-based} attack is the most straightforward one. It does not require calculating means and variances to form distributions or training a classification model. We can directly use the $\tau$ obtained from the shadow model to determine the membership of samples in the target model. Since the \textit{threshold-based} attack uses a one-dimensional threshold for judgment, it may reduce accuracy and become less stable to some extent. Therefore, before exploring other influence factors, we aim to validate whether the threshold obtained from the shadow model can be effectively used to attack the target model and to test the impact of different $\tau$ on attack accuracy.}

{From the results shown in~\autoref{fig:diff_threshold}, it can be observed that using the shadow model to determine the best threshold $\tau$ allows for a successful attack on the target model. The threshold $\tau$, calculated using sample data from the shadow model, is more effective at distinguishing between the target model's member and non-member samples compared to other nearby values. This consistent result is evident across the three datasets included in the experiment. The AUC of our four attacks all exceeds $0.7$, demonstrating the feasibility of threshold-based attacks. Moreover, the impact of different thresholds on attack performance is shown in~\autoref{fig:diff_threshold}. We found that even with a deviation (e.g., $0.02$), the $\tau$ obtained from the shadow model still achieves good attack accuracy on the target model.}

\begin{table*}[!t]
    \vspace{0.5cm}
    \belowrulesep=0pt
    \aboverulesep=0pt
    \centering
    \caption{{The AUC scores of three attack types (\textit{threshold-based}, \textit{distribution-based}, \textit{classifier-based}) across three datasets in four scenarios (\hyperref[Attack-I]{Attack-I}, \hyperref[Attack-II]{Attack-II}, \hyperref[Attack-III]{Attack-III}, \hyperref[Attack-IV]{Attack-IV}) highlighting Cosine similarity's superior and stable performance across all metrics and attack types. The best performance in each scenario is highlighted in bold.}}
    \label{fig:diff-metrix}
    \small
    \resizebox{0.80\textwidth}{!}{
    \begin{tabular}{cccccccccccccc}
    \toprule
         \multicolumn{2}{c}{\multirow{2}{*}{Method}}&\multicolumn{4}{c}{CelebA}&\multicolumn{4}{c}{WIT}&\multicolumn{4}{c}{WIT} \\ \cmidrule(lr){3-6} \cmidrule(lr){7-10} \cmidrule(lr){11-14} 
         & & $\ell_1$ & $\ell_2$ & Hamming & Cosine &$\ell_1$ & $\ell_2$ & Hamming & Cosine& $\ell_1$ & $\ell_2$ & Hamming & Cosine\\ \midrule
         \multirow{4}{*}{\text{\begin{tabular}[c]{@{}c@{}}\rotatebox[origin=c]{90}{\scriptsize Threshold} \end{tabular}}}& Attack-I &$0.30$ & $0.33$ & $0.71$ & $\boldsymbol{0.88}$ & $0.39$ & $0.38$ & $\boldsymbol{0.73}$ & $0.69$ & $0.40$ & $0.37$ & $0.78$ & $\boldsymbol{0.84}$\\
         & Attack-II & $0.27$ & $0.30$ & $0.69$ & $\boldsymbol{0.83}$ & $0.39$ & $0.39$ & $\boldsymbol{0.71}$ & $0.67$ & $0.44$ & $0.43$ & $0.80$ & $\boldsymbol{0.82}$ \\
         & Attack-III & $0.30$ & $0.34$ & $0.74$ & $\boldsymbol{0.82}$ & $0.40$ & $0.40$ & $0.67$ & $\boldsymbol{0.69}$ & $0.40$ & $0.39$ & $0.79$ & $\boldsymbol{0.84}$ \\
         & Attack-IV & $0.41$ & $0.43$ & $0.48$ & $\boldsymbol{0.82}$ & $0.43$ & $0.47$ & $\boldsymbol{0.77}$ & $0.66$ & $0.37$ & $0.38$ & $0.77$ & $\boldsymbol{0.74}$ \\ \hdashline[0.8pt/5pt]
         \multirow{4}{*}{\text{\begin{tabular}[c]{@{}c@{}}\rotatebox[origin=c]{90}{\scriptsize Distribution} \end{tabular}}}
            & Attack-I & $0.79$ & $0.83$ & $0.86$ & $\boldsymbol{0.93}$ & $0.82$ & $0.82$ & $0.79$ & $\boldsymbol{0.84}$ & $0.81$ & $0.81$ & $0.79$ & $\boldsymbol{0.82}$ \\
         & Attack-II & $0.83$ & $0.82$ & $0.83$ & $\boldsymbol{0.93}$ & $0.80$ & $\boldsymbol{0.83}$ & $0.77$ & $0.82$ & $\boldsymbol{0.82}$ & $0.81$ & $0.78$ & $0.81$ \\
         & Attack-III & $0.68$ & $0.67$ & $0.75$ & $\boldsymbol{0.94}$ & $0.67$ & $0.67$ & $0.56$ & $\boldsymbol{0.70}$ & $0.65$ & $0.66$ & $0.59$ & $\boldsymbol{0.72}$ \\
         & Attack-IV & $0.65$ & $0.66$ & $0.73$ & $\boldsymbol{0.88}$ & $0.66$ & $0.65$ & $0.68$ & $\boldsymbol{0.70}$ & $0.66$ & $0.66$ & $0.58$ & $\boldsymbol{0.67}$ \\ \hdashline[0.8pt/5pt]
         \multirow{4}{*}{\text{\begin{tabular}[c]{@{}c@{}}\rotatebox[origin=c]{90}{\scriptsize Classifier} \end{tabular}}} & Attack-I & $0.74$ & $0.84$ & $0.85$ & $\boldsymbol{0.93}$ & $0.73$ & $0.75$ & $0.78$ & $\boldsymbol{0.82}$ & $0.73$ & $0.77$ & $0.76$ & $\boldsymbol{0.86}$ \\
         & Attack-II & $0.79$ & $0.76$ & $0.86$ & $\boldsymbol{0.93}$ & $0.73$ & $0.73$ & $0.79$ & $\boldsymbol{0.82}$ & $0.77$ & $0.78$ & $0.74$ & $\boldsymbol{0.91}$ \\
         & Attack-III & $0.81$ & $0.75$ & $0.83$ & $\boldsymbol{0.94}$ & $0.70$ & $0.71$ & $0.78$ & $\boldsymbol{0.79}$ & $0.53$ & $0.73$ & $0.80$ & $\boldsymbol{0.83}$ \\
         & Attack-IV & $0.77$ & $0.73$ & $0.82$ & $\boldsymbol{0.93}$ & $0.75$ & $0.74$ & $0.70$ & $\boldsymbol{0.79}$ & $0.52$ & $0.62$ & $0.78$  & $\boldsymbol{0.82}$ \\
    \bottomrule
    \end{tabular}}
\end{table*}

\begin{figure*}[!t]
    \centering
    \includegraphics[width = 0.95\textwidth]{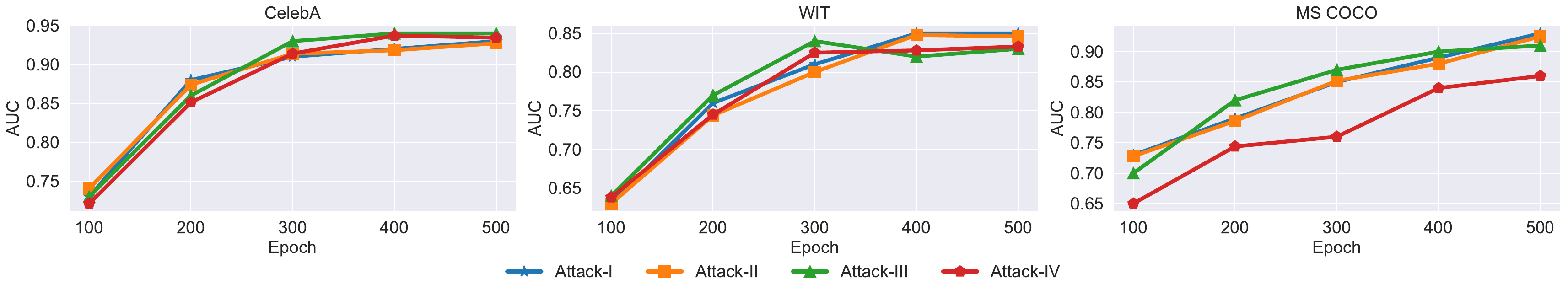}
    \caption{Relationship between epoch progression and AUC score in~\hyperref[Attack-I]{Attack-I},~\hyperref[Attack-II]{Attack-II},~\hyperref[Attack-III]{Attack-III}, and~\hyperref[Attack-IV]{Attack-IV}, indicating increasing memorization within image generation models over fine-tuning epochs.
}
    \label{fig:ft}
\end{figure*}

\subsection{Impact of Different Image Encoder} \label{sec:image_encoder}

As our attack is a similarity scores-based attack, and we measure the distance between the query image \( I_q \) and the image \( I_g \) generated by the target model using the embeddings \( E(I_g) \) and \( E(I_q) \). However, due to the multitude of high-performance image encoder models, each with its unique pre-trained dataset and model architecture, we employed five distinct image feature extractors: \texttt{DETR}~\cite{carion2020endtoend}, \texttt{BEiT}~\cite{bao2022beit}, \texttt{EfficientFormer}~\cite{li2022efficientformer}, \texttt{ViT}~\cite{dosovitskiy2021image}, and \texttt{DeiT}~\cite{touvron2021training}. Our goal was to observe the impact of various image features on the success rate of attacks by generating image embeddings from these models. The extractor yielding the highest success rate will be selected as the default image feature extractor for subsequent experiments.

As depicted in~\autoref{fg:encoder}, our five image feature extractors excel across four different attack scenarios within the \textit{classifier-based} attack domain. Each maintains an AUC score exceeding $0.7$, underscoring the robustness of our attack framework across different feature extractors. Notably, the implementation of \texttt{DeiT}~\cite{touvron2021training} as the feature extraction model yielded a marginally higher and more consistent success rate compared to the other image encoders. Therefore, we selected \texttt{DeiT} as the default image encoder for future experiments. 

A more comprehensive comparison including \textit{threshold-based} and \textit{distribution-based} of these five image encoders is presented in~\hyperref[appendix:image_encoder]{Appendix~\ref*{appendix:image_encoder}}.

\subsection{Impact of Different Distance Metrics}

In the previous section, we picked \texttt{DeiT}~\cite{touvron2021training} as the most stable and efficient image feature extractor. However, our attack framework also necessitates a reliable and consistent distance metric to compute the similarity score between embeddings. We conducted systematic and extensive experiments, and as demonstrated in~\autoref{fig:diff-metrix}, we thoroughly assessed various attack scenarios and types across all datasets to test their effects on Cosine similarity, $\ell_1$ distance, $\ell_2$ distance, and Hamming distance.

From~\autoref{fig:diff-metrix}, it is evident that using Cosine similarity as the distance metric yields optimal results for the computed distance vector, regardless of the attack scenario and type employed. We hypothesize that this phenomenon can be attributed to the focal point of our computation: the image embedding vectors generated by the encoder for both $I_q$ and $I_g$. Cosine similarity is inherently adept at measuring the similarity between two vectors. In contrast, $\ell_1$ and $\ell_2$ norms are more suitable for quantifying pixel-level discrepancies between $I_q$ and $I_g$, making them less efficient for evaluating the distance between two vectors.


\subsection{Impact of Fine-tuning Steps}
We then investigated the influence of the number of fine-tuning steps on the success rate of attacks. Evaluations were conducted at intervals of $100$ epochs, ranging from $100$ to $500$ epochs, to measure the attack success rate. {The default image encoder and distance metrics are \texttt{Deit} and Cosine similarity; all fine-tuning settings are aligned with~\autoref{tab:default_setting}.} As the model's memorization of the training data can be equated to overfitting effects, it is anticipated that with an increased number of fine-tuning steps, the model increasingly exhibits a tendency towards overfitting and enhanced memorization of the training samples. Consequently, when querying the model with member set samples compared to non-member samples, a more distinct similarity discrepancy should be observed.

In~\autoref{fig:ft}, we present the results of the \textit{classifier-based} attacks under four attack scenarios:~\hyperref[Attack-I]{Attack-I},~\hyperref[Attack-II]{Attack-II},~\hyperref[Attack-III]{Attack-III}, and~\hyperref[Attack-IV]{Attack-IV}. The outcomes indicate that~\hyperref[Attack-I]{Attack-I} and~\hyperref[Attack-III]{Attack-III} achieve higher success rates compared to the other two scenarios. This can be attributed to the fact that when utilizing the query data sample $x$, it inherently comprises the text caption $T_q$. As a result, neither~\hyperref[Attack-I]{Attack-I} nor~\hyperref[Attack-III]{Attack-III} require the employment of a caption model to generate corresponding text descriptions based on $I_q$. This circumvents the introduction of additional biases that could cause discrepancies between the model-generated images and $I_q$ itself.

We have also included the results for \textit{threshold-based} and \textit{distribution-based} attacks under these four scenarios in the~\hyperref[appendix:fine_steps]{Appendix~\ref*{appendix:fine_steps}} for reference.

\subsection{Impact of Number of Inference Step}

The quality of images generated by current diffusion models, including the Stable Diffusion~\cite{rombach2022high} presented in our work, is influenced not only by the number of fine-tuning steps but also by the number of inference steps. These models predominantly utilize DDIM~\cite{song2022denoising} as their sampling method. The {Fr\'{e}chet Inception Distance} is able to shift moderately from $13.36$ to $4.04$ when varying the sampling steps from $10$ to $1000$. This change highlights the capability of a higher number of inference steps to produce images of superior quality. Given that the foundation of our attack relies on the distance between generated and original images, we posit that an increased number of inference steps, which results in images closely resembling the original and of better quality, would correspondingly enhance the attack's success rate.

\begin{figure}[t]
    \centering
    \resizebox{0.45\textwidth}{!}{
    \includegraphics{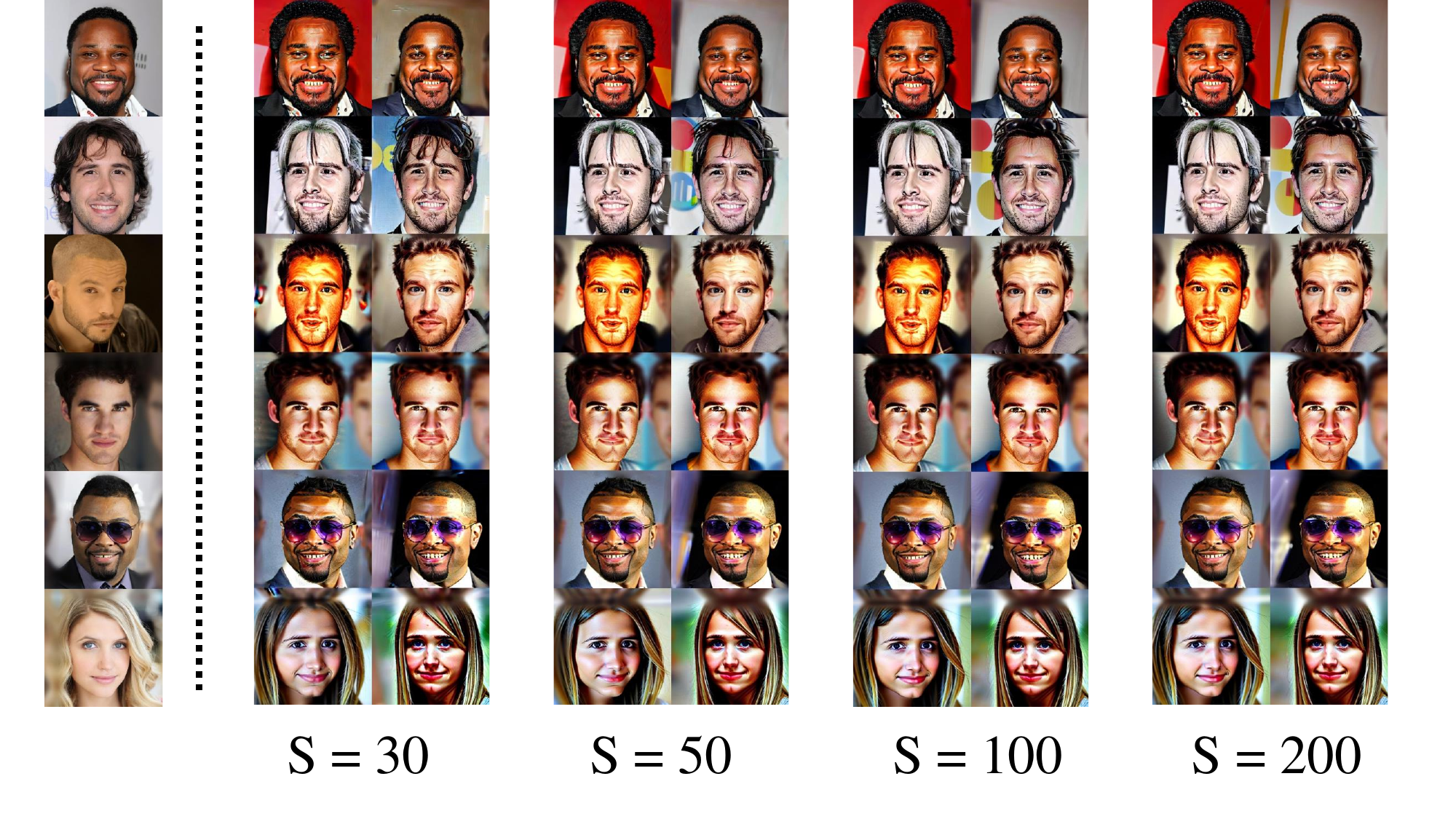}}
    \caption{{The inference steps of $30$, $50$, $100$, and $200$ showed no noticeable differences in the overall structure of the generated images. Only subtle details, such as hair, exhibited variations.}}
    \label{fig:inference_step}
\end{figure}

As illustrated in~\autoref{tab:inference_step}, the variations in attack accuracy are not immediately pronounced. However, upon a broader examination, it becomes evident that as the number of S (inference steps) increases, there is a gradual uptrend in the success rate of attacks. Notably, attacks based on classifiers yield the highest accuracy. To delve deeper into the reason why an increased number of inference steps does not lead to a substantial boost in attack success rate, we present samples generated at different inference steps in~\autoref{fig:inference_step}. It becomes apparent that as the number of inference steps rises, only certain localized features of the generated images are affected. The overall style remains largely undisturbed, with no significant discrepancies observed. This observation potentially explains why altering the inference steps does not drastically impact the attack success rate.

The experimental results obtained from the additional two datasets are presented in~\hyperref[appendix:inference_step]{Appendix~\ref*{appendix:inference_step}}.

\begin{table}[t]
    \caption{Alignment with DDIM~\cite{song2022denoising} denoting `S' as inference steps; experimentation under~\hyperref[Attack-III]{Attack-III} scenario measuring FID at varying inference step counts. For each type of attack, we highlight the optimal attack result corresponding to each evaluation metric.}
    \label{tab:inference_step}
    \centering
    \Huge
    \resizebox{0.48\textwidth}{!}{
    \begin{tabular}{c|cccccccccc}
        \toprule[1.1pt] 
        \multirow{2}{*}{S} & 
        \multicolumn{3}{c}{Threshold-based} & 
        \multicolumn{3}{c}{Distribution-based} &
        \multicolumn{3}{c}{Classifier-based} &
        \multirow{2}{*}{FID} \\
        \cmidrule[0.7pt](lr){2-4} \cmidrule[0.7pt](lr){5-7} \cmidrule[0.7pt](lr){8-10}
         & ASR & AUC & T@F=1\% & ASR & AUC & T@F=1\% & ASR & AUC & T@F=1\% \\ 
         \cmidrule[0.7pt](lr){1-10}
         $30$ & $\bm{0.75}$ & $\bm{0.8225}$ & $0.30$ & $0.76$ & $0.8816$ & $\bm{0.50}$ & $0.865$ & $0.93$  & $0.54$  &  $8.77$ \\
         $50$ & $0.765$ & $0.8146$ & $0.25$ & $\bm{0.77}$ & $\bm{0.8920}$ & $0.37$ & $0.85$ & $0.93$  & $0.58$  &  $7.835$ \\
         $100$ & $0.74$ & $0.8172$ & $0.26$ & $0.745$ & $0.8818$ & $0.40$ & $0.855$ & $\bm{0.94}$  & $\bm{0.61}$  &  $7.527$ \\
         $200$ & $0.745$ & $0.8125$ & $\bm{0.39}$ & $0.74$ & $0.8869$ & $0.49$ & $\bm{0.87}$ &  $\bm{0.94}$ &  $0.58$ & $7.472$\\ 
         \bottomrule[1.1pt]
    \end{tabular}
    }
\end{table}

\subsection{Impact of Different Size of Auxiliary Dataset}

From our observations across white-box~\cite{carlini2023extracting,matsumoto2023membership,hu2023membership}, gray-box~\cite{duan2023diffusion,kong2023efficient,hu2023membership}, and black-box attacks~\cite{matsumoto2023membership}, the accuracy of these attacks is significantly influenced by the size of training set. As the training set of the target model, encompasses more samples, its ``memorization'' capability for individual samples diminishes. This is attributed to the fact that an increase in training data can decelerate the model's convergence rate, impacting its ability to fit all the training sets accurately. As a result, many attacks do not demonstrate effective performance as the dataset size expands. In this work, we investigate how increasing the size of the dataset used by the target model affects the success rate of our black-box attack. Given that our work is predicated on leveraging pre-trained models for downstream tasks, where the downstream datasets usually do not contain a vast number of samples, we have established our training dataset sizes at $100$, $200$, $500$, and $1000$. Using the CelebA dataset, we aim to assess the variations in the performance of the three attack types when the attacker is privy to four distinct values of knowledge.

As illustrated in~\autoref{fig:size}, the attack success rate tends to decrease as the number of images in the training set increases. However, even when the users use $1,000$ samples to fine-tune the target models, in the scenarios of \hyperref[Attack-I]{Attack-I} and \hyperref[Attack-III]{Attack-III}, a classifier used as the attack model can still achieve a success rate of over $60\%$.

\subsection{Impact of the Selection of Shadow Models}
To examine the generalization and applicability of our attack methodology in real-world scenarios, we propose to further relax the assumptions pretraining to the attack environment. In our prior experiments, all results were predicated on the use of shadow models mirroring the target model's structural framework to generate training data for the attack inference model. However, in practical settings, malicious model publishers may withhold any specific details about the model, offering only a user interface. Under such circumstances, it is not advisable to confine ourselves to a specific type of shadow model. Instead, a more effective approach would be to leverage the memorization properties of image generators when creating training data for the attack, thus diversifying and strengthening the attack strategy.

Therefore, we employed a conditional image generator, Kandinsky~\cite{razzhigaev2023kandinsky}, which has a different architectural design from Stable Diffusion~\cite{rombach2022high}, as our shadow model. This model was fine-tuned using the same auxiliary dataset mentioned in~\autoref{tab:default_setting}, and the results are displayed in~\autoref{tab:ablation}. 


\begin{table}[t]
    \belowrulesep=0pt
    \aboverulesep=0pt
    \caption{Use of Kandinsky~\cite{razzhigaev2023kandinsky} as shadow model and Stable Diffusion~\cite{rombach2022high} as target model in conducting attacks, demonstrating the maintained efficacy of all four attack scenarios. We use the `X'-`Y' format to represent different experiment settings in the table, where `X' means four attack scenarios, `Y' being `S' or `A' denotes whether the shadow model is the same as or different from the target model. For each comparison, the optimal result is marked in bold.}
    \label{tab:ablation}
    \centering
    \resizebox{0.48\textwidth}{!}{
    \begin{tabular}{c|cc|cc|cc|cc}
        \toprule[1.1pt]
        \multicolumn{1}{c|}{Dataset} & I-S & I-A & II-S & II-A & III-S & III-A & IV-S & IV-A  \\
        \hline
        CelebA  & $\bm{0.93}$ & $0.87$ & $\bm{0.93}$ & $0.86$ & $\bm{0.93}$ & $0.86$ & $\bm{0.93}$ & $0.85$ \\
        WIT  & $\bm{0.83}$ & $0.81$ & $0.83$ & $\bm{0.84}$ & $\bm{0.84}$ & $\bm{0.84}$ & $\bm{0.83}$ & $\bm{0.83}$ \\
        MS COCO  & $\bm{0.92}$  & $0.89$ & $\bm{0.92}$ & $0.91$ & $\bm{0.89}$ & $\bm{0.89}$ & $\bm{0.76}$ & $0.74$ \\
        \bottomrule[1.1pt]
    \end{tabular}}
\end{table}

In~\autoref{tab:ablation}, we evaluate attackers with different knowledge across three datasets, employing a classifier as the attack inference model. The notation `\raisebox{-0.4ex}{*}-\raisebox{-0.4ex}{*}-S' indicates attacks conducted using a shadow model with the same architecture as the target model. Conversely, `\raisebox{-0.4ex}{*}-\raisebox{-0.4ex}{*}-A' denotes scenarios where the target model is anonymous to the attacker, hence the shadow model and the target model are architecturally dissimilar. The experimental data indicate that altering the shadow model has only a minimal effect on the success rate of the attacks, with all attacks still capable of achieving a relatively high level of success. This further substantiates the robustness and generalizability of our attack framework.

\subsection{Impact of Eliminating Fine-Tuning in Captioning Models} \label{fine-tuning-caption}
In our work, within the attack environments designed for~\hyperref[Attack-II]{Attack-II} and~\hyperref[Attack-IV]{Attack-IV}, the attacker does not have full access to the query point $x$, but only a query image $I_q$. In previous sections, for these two attack scenarios, we initially used auxiliary data to fine-tune the image captioning model before generating matching prompt information based on the query image. However, this approach significantly increases the time cost of the attack. Therefore, we use an image captioning model that has not been fine-tuned to generate image descriptions. We then carry out the attack based on these generated descriptions.

From~\autoref{tab:unfine}, it is evident that without fine-tuning the captioning model, there is a varying degree of reduction in the success rates of attacks across different datasets. Notably, when using CelebA-Dialog as the test set, the success rate of the attack drops by nearly $30\%$, leading to a marked inconsistency in the attack outcomes. Unlike changing the types of shadow models, a captioning model without tuning more conspicuously diminishes the effectiveness of the attacks. We posit the image captioning model may have introduced biases in the generated {text component} $T_q$, adversely affecting the quality of the resultant images.

\begin{tcolorbox}[breakable, colback=takeaways, boxrule=0pt]
\textit{Takeaways:}\: We compared the four attack scenarios we proposed with existing black-box attacks and found that our accuracy significantly surpasses the established baselines. Then, we tested the feasibility of implementing the most straightforward threshold-based attack using compressed high-dimensional similarity scores. To thoroughly evaluate the accuracy and stability of our attacks, we conducted tests employing various image encoders, distance metrics, fine-tuning steps, and inference procedures, as well as different sizes of auxiliary datasets. Additionally, we experimented with changing the types of shadow models and testing without fine-tuning the image caption model to test our attacks' generalization and robustness. Our findings reveal a strong correlation between the attacks' success rate and the generated images' quality. Higher quality images lead to increased attack success rates, which aligns with the theory of similarity scores mentioned in~\autoref{theoretical_foundation}. 
\end{tcolorbox}

\begin{table}[t]
    \vspace{0.2cm}
    \caption{Impact of not fine-tuning the captioning model on the success rates of~\hyperref[Attack-II]{Attack-II} and~\hyperref[Attack-IV]{Attack-IV} across various datasets. The best result for each scenario is marked in bold.}
    \label{tab:unfine}
    \centering
    \Huge
    \resizebox{0.405\textwidth}{!}{
    \begin{tabular}{ccccc}
        \toprule[1.1pt]
         \multirow{2}{*}{Dataset} & \multicolumn{2}{c}{Attack-II} & \multicolumn{2}{c}{Attack-IV} \\
         \cmidrule[0.7pt](lr){2-3} \cmidrule[0.7pt](lr){4-5}
          & With tuning & W/o tuning & With tuning & W/o tuning\\
          \cmidrule[0.7pt](lr){1-5}
          CelebA & $\bm{0.93}$ & $0.59$ & $\bm{0.93}$ & $0.60$\\
          WIT & $\bm{0.83}$ & $0.70$ & $\bm{0.83}$ & $0.56$\\
          MS COCO & $\bm{0.93}$ & $0.79$ & $\bm{0.73}$ & $0.65$ \\
         \bottomrule[1.1pt]
    \end{tabular}}
\end{table}

\begin{table*}

    \caption{{Attack accuracy under DP-SGD defense. Our four attack methods' accuracy declines. Experiments include two different sizes of datasets and three $\epsilon$ values. `Vanilla' means without DP-SGD. The highest accuracy is marked in bold.}}
    \label{tab:defense}
    \centering
    \resizebox{0.80\textwidth}{!}{
    \begin{tabular}{c|ccccccccccccc}
    \toprule[1.1pt]
         \multicolumn{2}{c}{} & \multicolumn{3}{c}{\textbf{Attack-I}}&\multicolumn{3}{c}{\textbf{Attack-II}}&\multicolumn{3}{c}{\textbf{Attack-III}}&\multicolumn{3}{c}{\textbf{Attack-IV}}\\ \cmidrule[0.7pt](lr){3-5} \cmidrule[0.7pt](lr){6-8} \cmidrule[0.7pt](lr){9-11} \cmidrule[0.7pt](lr){12-14}
         & &ASR$^{\uparrow}$&AUC$^{\uparrow}$ &T@1\%F$^{\uparrow}$&ASR$^{\uparrow}$&AUC$^{\uparrow}$&T@1\%F$^{\uparrow}$&ASR$^{\uparrow}$&AUC$^{\uparrow}$&T@1\%F$^{\uparrow}$&ASR$^{\uparrow}$&AUC$^{\uparrow}$&T@1\%F$^{\uparrow}$\\ \cmidrule[0.7pt](lr){1-14}
         \multirow{4}{*}{\textbf{\begin{tabular}[c]{@{}c@{}} $100$ \end{tabular}}}&$\epsilon = 1$&$0.581$&$0.646$&$0.01$&$0.532$&$0.654$&$0.01$&$0.495$&$0.498$&$0.00$&$0.522$&$0.524$&$0.00$\\ 
         &$\epsilon = 4$&$0.592$&$0.651$&$0.01$&$0.575$&$0.647$&$0.01$&$0.515$&$0.514$&$0.01$&$0.535$&$0.534$&$0.01$\\ 
         &$\epsilon=10$&$0.595$&$0.641$&$0.02$&$0.560$&$0.644$&$0.02$&$0.56$&$0.522$&$0.01$&$0.545$&$0.522$&$0.01$\\ &Vanilla&$\bm{0.843}$&$\bm{0.911}$&$\bm{0.58}$&$\bm{0.845}$&$\bm{0.909}$&$\bm{0.51}$&$\bm{0.831}$&$\bm{0.893}$&$\bm{0.38}$&$\bm{0.765}$&$\bm{0.813}$&$\bm{0.19}$\\ \cmidrule[0.7pt](lr){1-14}
         \multirow{4}{*}{\textbf{\begin{tabular}[c]{@{}c@{}} $200$ \end{tabular}}}&$\epsilon = 1$&$0.593$&$0.632$&$0.01$&$0.628$&$0.676$&$0.01$&$0.493$&$0.502$&$0.00$&$0.548$&$0.524$&$0.01$\\
         &$\epsilon = 4$&$0.601$&$0.652$&$0.01$&$0.618$&$0.670$&$0.01$&$0.523$&$0.516$&$0.01$&$0.515$&$0.506$&$0.01$\\
         &$\epsilon = 10$&$0.585$&$0.632$&$0.03$&$0.643$&$0.655$&$0.02$&$0.535$&$0.504$&$0.01$&$0.542$&$0.541$&$0.02$\\
         &Vanilla&$\bm{0.767}$&$\bm{0.863}$&$\bm{0.30}$&$\bm{0.730}$&$\bm{0.812}$&$\bm{0.11}$&$\bm{0.695}$&$\bm{0.728}$&$\bm{0.09}$&$\bm{0.773}$&$\bm{0.800}$&$\bm{0.14}$\\
    \bottomrule[1.1pt]
    \end{tabular}}
\end{table*}

\section{Defense} \label{defense}


{
In this part, we want to employ Differential Privacy Stochastic Gradient Descent (DP-SGD)~\cite{abadi2016deep} to evaluate the robustness of our attack. DP-SGD adds noise into the gradient during the training phase and provides a guarantee that the presence/absence of any single training sample only incurs quantifiably limited differences~\cite{dwork2016calibrating} and thus diminishes the model's memorization of individual samples. 
}

We tested our four attacks and employed a classifier as the inference model against the MS COCO dataset. {Due to the limit of time and computing resources, we only selected two different sizes of datasets, $100$ and $200$. For the DP-SGD mechanism, we set clipping norm $C=1$, $\delta=1\times10^{-3}$, sampling rate $q=4/$(dataset size), epoch number is $500$, and target privacy budget (with a slight abuse of notation) $\epsilon \in \{1,4,10\}$ (different $\epsilon$ gives different noise multiplier $\sigma$). }


{The experimental results can be seen in~\autoref{tab:defense}. It illustrates that the attack success rate of four attacks significantly decreases after implementing DP-SGD~\cite{abadi2016deep} as the defensive method. When we use $100$ samples to fine-tune the model and set $\epsilon=1$, four attacks have been greatly impacted dropping to around $50\%$ (random guess). When we change the $\epsilon$ value from $1$ to $4$ and $10$, the attack accuracy increases but still cannot show their effectiveness. From~\autoref{tab:defense}, we noticed TPR at FPR=$1\%$ has dropped to $0.01$. This phoneme further demonstrates that~\hyperref[Attack-I]{Attack-I},~\hyperref[Attack-II]{Attack-II},~\hyperref[Attack-III]{Attack-III}, and~\hyperref[Attack-IV]{Attack-IV} all lose their functionality in these defense settings.}

\begin{figure}[t]
    \centering
    \includegraphics[width=0.90\linewidth]{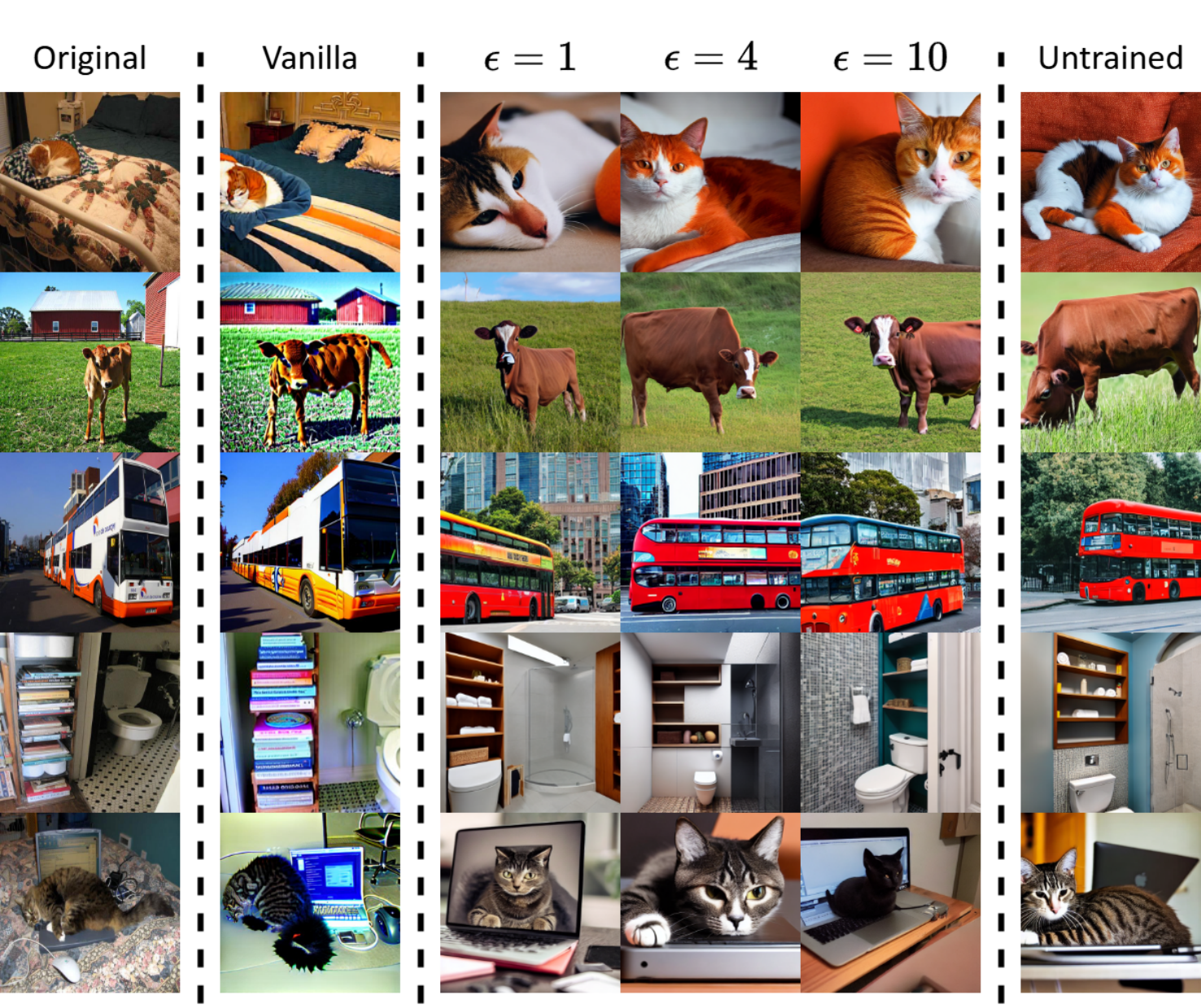}
    \caption{{Effect of adding DP-SGD~\cite{abadi2016deep} on model memorization. `Original' represents the training samples, while `Vanilla' denotes samples generated after fine-tuning without using DP-SGD. $\epsilon = 1, 4, 10$ indicate samples generated by the fine-tuned model after applying DP-SGD at varying levels. `Untrained' represents samples generated by the Stable Diffusion v1-5 without fine-tuning. The generated images in the same row are from the same prompt.}}
    \label{fig:defense_img}
\end{figure}

{To understand how DP-SGD mitigates our attack, we presented the generated images from different settings in~\autoref{fig:defense_img}. We observe that compared to the `Vanilla' version (fine-tune without DP-SGD), DP-SGD prevents the model's memorization of training samples. For instance, in the first row, it is clear that DP-SGD omits detailed features of the training images, such as pillows. This effect is observed regardless of the value of $\epsilon$. The generated images with defense remain very similar to those produced by the untrained Stable Diffusion v1-5 model. DP-SGD weakens the model's memorization of the training samples, thereby reducing the similarity score and rendering our four attacks ineffective.}






\section{Related Work} \label{related_work}
We further review related work on white-box and gray-box membership inference attacks against diffusion models. 

\subsection{White-Box MIA}
In the white-box setting, the attacker has access to the parameters of the victim model. Note that in MIA for classification tasks, it is observed that having black-box means can sufficient enough information (e.g., predict vector~\cite{chen2021machine, hui2021practical, jayaraman2021revisiting, long2018understanding, shokri2021privacy, shokri2017membership}, top-$k$ confidence score~\cite{salem2018mlleaks,shokri2017membership}); but in MIA for generative models, because the model is more complicated and directly applying existing MIAs is not successful, white-box attacks are investigated.  

Both Hu et al.~\cite{hu2023membership} and Matsumoto et al.~\cite{matsumoto2023membership} adopt approaches similar to that of Yeom et al.~\cite{yeom2018privacy}, determining membership by comparing the loss at various timesteps to a specific threshold. Carlini et al.~\cite{carlini2023extracting} argue that mere threshold-based determinations are insufficient and proposed training multiple shadow models and utilizing the distribution of loss across each timestep established by these shadow models to execute an online LiRA attack~\cite{carlini2022membership}. Pang et al.~\cite{pang2023white} leveraged the norm of gradient information computed from timesteps uniformly sampled across total diffusion steps as attack data to train their attack model.

\subsection{Gray-box MIA}
Gray-box access does not acquire any internal information from the model. However, given that diffusion models generate images through a progressive denoising process, attacks in this setting assume the availability of intermediate outputs during this process. {In particular, several works leveraged the deterministic properties of generative process in DDIM~\cite{song2022denoising} for their attack designs.} Duan et al.~\cite{duan2023diffusion} employed the approximated posterior estimation error as attack features, while Kong et al.~\cite{kong2023efficient} used the magnitude difference $\lVert x_{t-t'} - x'_{t-t'}\rVert_p$ from the denoising process as their attack criterion, where $x_{t-t'}$ represents the ground truth and $x'_{t-t'}$ denotes the predicted value. Fu et al.~\cite{fu2023probabilistic} use the intermediate output to calculate the probabilistic fluctuations between target points and neighboring points. {Similarly, Zhai et al.~\cite{zhai2024membership} sampled multiple times at different denoising steps, with the likelihood discrepancy between the conditional and unconditional generations as the criterion. Fu et al.~\cite{fu2024model} based their approach on the structural similarity between intermediate outputs and the original images. Li et al.~\cite{li2024unveiling} found that the similarity between reconstructed images and the original images after degradation can also serve as a standard for evaluation.
}

\section{Conclusion} \label{conclusion}


In this work, we introduce a black-box membership inference attack framework specifically designed for contemporary conditional diffusion models. Given the rapid development of diffusion models and the abundance of open-source pre-trained models available online, we focus on the potential privacy issues arising from utilizing these pre-trained models fine-tuned for downstream tasks. Recognizing the absence of effective attacks against the current generation of conditional image generators, we leverage the objective function of diffusion models to propose a black-box similarity scores-based membership inference attack. Our experiments not only demonstrate the flexibility and effectiveness of this attack but also highlight significant privacy vulnerabilities in image generators, underscoring the need for increased attention to these issues.

However, our attacks still face certain limitations. As discussed in~\autoref{fine-tuning-caption}, both~\hyperref[Attack-II]{Attack-II} and~\hyperref[Attack-IV]{Attack-IV} critically rely on a captioning model that has been fine-tuned using an auxiliary dataset. We hope future work can effectively address this challenge.

\section*{Acknowledgement}

We thank the shepherd and the reviewers for their valuable comments and suggestions. This paper is fully supported by CCF-2217071 and OAC-2319988.






\bibliographystyle{IEEEtran}
\bibliography{bib}

\appendices

\section{More Details for Diffusion Models}\label{appendix:diffusion}

Diffusion models have two phases: the forward diffusion process and the reverse denoising process. In the forward process, an image $x$ is sampled from the true data distribution. The image $x$ undergoes a series of noise addition steps for $T$ iterations, resulting in a sequence $x_1, x_2, \ldots, x_{T-1}, x_T$, until $x_T$ becomes an image equation to an isotropic Gaussian noise distribution. The magnitude of noise introduced at each step is controlled by a parameter $\alpha_t$, where $\alpha_t \in [0, 1]$, which gradually decreases over time.
At step $t$, the noise image $x_t$ can be represent as:
\begin{equation*}
    x_t = \sqrt{\alpha_t}x_{t-1} + \sqrt{1-\alpha_t} \epsilon; \quad\epsilon \sim \mathcal{N}(0,1)
\end{equation*}
Given that the original work~\cite{ho2020denoising} considers the forward process as a Markov chain and employs the reparameterization trick, it is possible to directly derive the noisy image $x_t$ at step $t$ from the original image $x_0$.
Based on the definition $\bar{\alpha}_t = \prod_{i=1}^{t} \alpha_i$, the training objective for diffusion models is to obtain a denoising network capable of sampling $x_{t-1}$ from $x_t$ according to the distribution $\mathcal{N}(x_{t-1}; \mu_{\theta}(x_t, t), \sigma_t^2 \text{I})$. When conditioned on $x_0$ and $x_t$ from $q(x_{t-1} | x_t,x_0)$, the ground-truth distribution of $x_{t-1}$ is given by $\mathcal{N}(x_{t-1}; \tilde{\mu}_t(x_t, x_0), \sigma_t^2 \text{I})$. Given that variance is fixed as a hyperparameter, the focus is on calculating the difference between $\mu_{\theta}(x_t, t)$ and $\tilde{\mu}_t(x_t, x_0)$. Applying Bayes' rule to the ground-truth distribution 
\begin{equation}
    \tilde{\mu}_t(x_t, x_0) = \frac{\sqrt{\alpha_t}(1-\bar{\alpha}_{t-1})}{1-\bar{\alpha}_t}x_t + \frac{\sqrt{\bar{\alpha}_{t-1}}(1-\alpha_t)}{1-\bar{\alpha}_t}x_0 \label{eq:mu_t}
\end{equation}

The objective of the training process is to closely approximate $\mu_{\theta}(x_t, t)$ with $\tilde{\mu}_t(x_t, x_0)$. Then, parameterize 
\begin{equation}
    \mu_{\theta}(x_t, t) = \frac{\sqrt{\alpha_t}(1-\bar{\alpha}_{t-1})}{1-\bar{\alpha}_t}x_t + \frac{\sqrt{\bar{\alpha}_{t-1}}(1-\alpha_t)}{1-\bar{\alpha}_t}\hat{x}_\theta(x_t,t) \label{eq:mu_theta}
\end{equation}


\section{More Details for Classifier-free Guidance} \label{appendix: class-free}

As the field advances, diffusion models can create content from user-given prompts, primarily using classifier-free guidance~\cite{ho2022classifierfree}. Many diffusion models, including Imagen~\cite{saharia2022photorealistic}, DALL·E 2~\cite{ramesh2022hierarchical}, and Stable Diffusion~\cite{rombach2022high}, which utilize the classifier-free guidance mechanism, are trained on dual objectives; however, they can be represented with a single model during training by probabilistically setting the conditional prompt to null. A conditional generation without an explicit classifier is achieved using the denoising network $\bar{\mathcal{U}}_{\theta}(x_t, t, p)$, where 
\begin{equation*}
    \bar{\mathcal{U}}_{\theta}(x_t, t, p) = (w+1)\cdot\mathcal{U}_{\theta}(x_t, t, p) - w\cdot\mathcal{U}_{\theta}(x_t, t).
\end{equation*}

\noindent The variable $w$ represents the guidance scale factor, where a higher value of $w$ results in improved alignment between image and text at the potential expense of image fidelity.

\section{More Details for Theoretical Foundation}\label{appendix:Theorem}

\subsection{Proof for Theorem 1}\label{appendix:theorem1_proof}

\begin{proof}
    Diffusion models employ the ELBO to approximate the log-likelihood $p(x)$ of the entire training dataset.
\begin{align}
    \log p({x}) &\geq \mathbb{E}_{q(x_{1:T} | x_0)} \left[ \log \frac{p(x_{0:T})}{q(x_{1:T}|x_0)} \right] \nonumber \\
    &\cdots \nonumber \\ 
    &= \underbrace{\mathbb{E}_{q(x_1|x_0)}\left[ \log p_{\theta}(x_0|x_1) \right]}_{L_0} - \underbrace{\mathcal{D}_{KL}(q(x_T|x_0) \| p(x_T))}_{L_T} \nonumber \\
    &- \underbrace{\sum_{t=2}^{T}\mathbb{E}_q(x_t|x_0)\left[ \mathbf{D}_{KL}(q(x_{t-1}|x_t,x_0) \| p_{\theta}(x_{t-1}|x_t)) \right]}_{L_{t-1}}
    \label{eq:likelihood}
\end{align}

The primary focus of optimization is on $L_{t-1}$, as explicated in the original work~\cite{ho2020denoising}. The other terms are treated as constants and independent decoders. {The objective function can be rewritten as:} 
\[ \min \mathbf{D}_{KL}(q(x_{t-1}|x_t,x_0)\;  \| \; p_{\theta}(x_{t-1}|x_t)). \]
Based on the assumption in DDPM~\cite{ho2020denoising}, to elucidate further:
\begin{flalign}
&\underset{\theta}{\mathrm{arg\,min}} \, \mathbf{D}_{KL}(q(x_{t-1}|x_t,x_0)\; \| \; p_{\theta}(x_{t-1}|x_t)) \nonumber & \\ 
= &\underset{\theta}{\mathrm{arg\,min}} \mathbf{D}_{KL}(\mathbf{N}(\tilde{\mu}_{t}(x_t,x_0),\sigma_t^2\textbf{I}) \; \| \; \mathbf{N}(\mu_{\theta}(x_t,t),\sigma_t^2\textbf{I})) & 
\nonumber \\
= &\underset{\theta}{\mathrm{arg\,min}} \frac{1}{2\sigma_t^2} \left[ \left\| \tilde{\mu}_{t}(x_t,x_0) -  \mu_{\theta}(x_t,t) \right\|^2_2 \right] \label{eq:delta_mu} &
\end{flalign}
\quad In~\autoref{eq:delta_mu}, $q(x_{t-1}|x_t, x_0)$ represents the ground truth distribution of $x_{t-1}$ given $x_t$ and $x_0$, while $p_{\theta}(x_{t-1}|x_t)$ denotes the predicted distribution of $x_{t-1}$ parameterized by $\theta$. The term $\tilde{\mu}_{t}(x_t, x_0)$ corresponds to the mean of the ground truth distribution $q(x_{t-1}|x_t, x_0)$, and $\mu_{\theta}(x_t, t)$ corresponds to the mean of the predicted distribution $p_{\theta}(x_{t-1}|x_t)$.

From~\autoref{eq:mu_t} and~\autoref{eq:mu_theta} in~\hyperref[appendix:diffusion]{Appendix~\ref*{appendix:diffusion}} (which gives more details about diffusion models), we can rewrite~\autoref{eq:delta_mu} as:
\begin{equation}
    \underset{\theta}{\mathrm{arg\,min}} \frac{1}{2\sigma_t^2} \frac{\bar{\alpha}_{t-1}(1-\alpha_t)^2}{(1 - \bar{\alpha}_t)^2} \left[ \left\| x_0 - \hat{x}_{\theta}(x_t,t) \right\|^2_2 \right] \label{eq:delta_x}
\end{equation}
\quad ~\autoref{eq:delta_x} can also be further developed by substituting and expressing $x_0$ using $x_t$ according to~\autoref{eq:x_t}, and by introducing $\epsilon_t$ as the targeted prediction of the diffusion model, aligning with the optimization objectives stated in both DDPM~\cite{ho2020denoising} and DDIM~\cite{song2022denoising}. However, our aim is to demonstrate that the optimization goal of the diffusion model supports the use of similarity scores as an indicator for determining the membership of query data. Consequently, the objective function is merely reformulated in the form of~\autoref{eq:delta_x}. Given that the likelihood of all training data should be higher than that of data not in the training set, and as inferred from~\autoref{eq:likelihood} and~\autoref{eq:delta_x}, if a data point $x$ has a higher likelihood, the norm $\| x_0 - \hat{x}_{\theta}(x_t,t) \|$ at any timestep in the model should be smaller, indicating that the image generated by the model is closer to the original image. This can be expressed as:

\begin{flalign}
    \Pr{b=1| x, \theta} \propto -{\left\|  x_0 - \hat{x}_{\theta}(x_t,t)\right\|^2_2} \label{eq:pro}
\end{flalign}

\end{proof}

\subsection{Proof for Theorem 2}\label{appendix:theorem2_proof}





\begin{proof}
    In the original paper~\cite{rombach2022high}, the loss function of the Stable Diffusion model is described as follows:
    \begin{equation*}
        L_{LDM} = \mathbb{E}_{\mathcal{E}(x),\epsilon \sim \mathcal{N}(0,1),t} \left[ \| \epsilon_t - \mathcal{U}_{\theta}(z_t,t,\phi_{\theta}(p)) \|^2_2 \right]
    \end{equation*}
    The latent code $z_t$ is of a much smaller dimension than that of the original image. The denoising network $\mathcal{U}_{\theta}$ predicts the noise at timestep $t$ based on $z_t$ and the embedding generated by $\phi_{\theta}$, which takes $p$ as its input. Given that the forward process of the Stable Diffusion~\cite{rombach2022high} is fixed,~\autoref{eq:x_t} remains applicable. Therefore, by substituting in the expression $ \epsilon_t = \frac{z_t - \sqrt{\bar{\alpha_t}}z_0}{\sqrt{1-\bar{\alpha}_t}} $ and discarding other weight terms, we can rederive the loss function of the Stable Diffusion model as:
    \begin{equation}
        L_{LDM} = \mathbb{E}_{\mathcal{E}(x),t} \left[ \| z_0 - \hat{z}_{\theta}(z_t,t,\phi_{\theta}(p)) \|^2_2 \right]
        \label{eq:stable_loss}
    \end{equation}
    \quad As seen from~\autoref{eq:stable_loss}, Stable Diffusion is essentially trained to optimize image predictions at any given timestep to closely approximate the original image $\mathrm{D}(z_0)$, where $\mathrm{D}$ is the decoder in Stable Diffusion. For the Stable Diffusion model, we can still distinguish between member samples and non-member samples by the similarity scores $\left\| \mathrm{D}(z_0) - \mathrm{D}(\hat{z}_{\theta}(z_t, t, \phi_{\theta}(p))) \right\|^2_2$, which is expressed as:
    \begin{equation}
        \Pr{b=1| x, \theta} \propto -{\left\| \mathrm{D}(z_0) - \mathrm{D}(\hat{z}_{\theta}(z_t, t, \phi_{\theta}(p)))\right\|^2_2} \label{eq: ini_stable}
    \end{equation}
    
\end{proof}

\section{More Details for Traditional Black-box Attacks}  \label{appendix: discussion}

\paragraph{Monte Carlo Attack.} Hilprecht et al.~\cite{hilprecht2019monte} argue that generative models can overfit and memorize data due to their ability to capture specific details. Given a query sample $x$, attackers can utilize the generative model to sample $k$ images. Define an $\epsilon$-neighborhood set $U_{\epsilon}(x)$ as $U_{\epsilon}(x) = \{ x' \mid d(x, x') \leq \epsilon \}$. Intuitively, if a larger number of $g_i$ are close to $x$, the probability $\Pr{x' \in U_{\epsilon}(x)}$ will also be greater.
Through the Monte Carlo Integration~\cite{owen2013monte}, the Monte Carlo attack can be expressed as:
\begin{equation}
    \hat{f}_{MC-\epsilon}(x) = \frac{1}{k}\sum_{i=1}^{k}\mathbbm{1}_{x'_{i}\in U_{\epsilon}(x)}\label{equ:MC}
\end{equation}
Furthermore, they try to employ the Kernel Density Estimator (KDE)~\cite{parzen1962estimation} as a substitute for $\hat{f}_{MC-\epsilon}(x)$. The estimation of the likelihood of $x$ using KDE can be expressed as:
\begin{equation*}
    \hat{f}_{KDE}(x) = \frac{1}{nh^{d}} \sum^{k}_{i=1}K\left( \frac{x - x'_i}{h^{d}} \right) 
\end{equation*}
where $h_d$ is the bandwidth, {$n$ denotes number of samples generated from generator}, and $K$ is the Gaussian kernel function. If $x \in \mathcal{D}_m$, the likelihood $\hat{f}_{KDE}(x)$ should be substantially higher than the likelihood when $x \in \mathcal{D}_{nm}$. However, upon experimentation, it was observed that using $\hat{f}_{KDE}(x)$ as the criterion for the attack did not yield attack accuracy better than random guessing.

\paragraph{GAN-Leaks Attack.} In a further examination of the memorization with generative models, Chen et al.~\cite{chen2020gan} posited that the closer the generated data distribution $p_{\theta}(\hat{x})$ is to the training data distribution $q(x)$, the more likely it is for $\mathcal{G}$ to generate a query datapoint $x$. They further articulated this observation as: 
\begin{equation*}
    \Pr{y_q = 1 | x, \theta} \propto \mathsf{Pr}_{\mathcal{G}}[x|\theta_v]
\end{equation*}
However, due to the inability to represent generated data distribution with an explicit density function, computing the precise probability becomes intractable. Therefore, Chen et al.~\cite{chen2020gan} also employed the KDE method~\cite{parzen1962estimation} and sampled $k$ times to estimate the likelihood of $x$. This can be expressed as:
\begin{equation}
    \mathsf{Pr}_{\mathcal{G}}(x|\theta) = \frac{1}{k} \sum_{i=1}^{k} K(x, \mathcal{G}(z_i)); \quad z_i \sim P_z \label{equ:gan_KDE}
\end{equation}
Here, $K$ denotes the kernel function, and $z_i$ represents the input to $\mathcal{G}$, which sample from latent code distribution $P_z$. Besides, Chen et al.~\cite{chen2020gan} propose the approximation of~\autoref{equ:gan_KDE}:
\begin{equation}
   \mathsf{Pr}_{\mathcal{G}}(x|\theta_v) \approx \frac{1}{k} \sum_{i=1}^{k} \text{exp}(-d(x, \mathcal{G}(z_i))); \quad z_i \sim P_z \label{equ:gan_approximate}
\end{equation}
%
For the $k$ samples from $\mathcal{G}$, we use the distance metric $d(\cdot,\cdot)$ to measure and sum the distances between each sample $g_i$ and the query point $x$.~\autoref{equ:MC} and~\autoref{equ:gan_approximate} indicate that the only practical way to improve attack success rates in models with such stochastic sampling is by significantly increasing the number of samples $g_i$. For a full-black attack, around $100k$ samples are required for a query point $x$ to achieve an attack AUC close to $0.60$. This undoubtedly results in significant overhead. Then, Chen et al.~\cite{chen2020gan} introduced the concept of a partial-black attack. Specifically, the attacker first employs 
\[ z^\ast = \underset{z}{\mathrm{arg\,min}} L(x, \mathcal{G}(z)) \]
to identify the optimal latent code $z^\ast$. Subsequently, $g_i$ is generated using $\mathcal{G}(z^\ast)$ to compare with $x$. The partial-black attack method boosts the success rate of attacks and uses fewer data samples, but requires finding the optimal $z^\ast$. In GANs, the input latent code $z$ is typically a 100-dimensional random noise. However, in newer conditional diffusion models, the complexity and size of input embeddings have greatly increased. Additionally, the latest diffusion models use explicit prompt information instead of random latent codes, making the partial-black attack strategy used for GANs unsuitable for these new-generation generative models.

\section{More Details on the Attack Framework} \label{appendix:framework}

{We use an example to show how our attack works. Assume we have a query image $I_q$ and a generated image $I_g$. After extracting features using an image feature extractor $E$, the resulting vector has dimensions $[\text{patch\_size}, \text{latent\_size}]$. For example, when using \texttt{ViT} as $E$, the height and width of the image are first resized to $224\times 224$, and then divided into $196$ patches, each of size $16 \times 16$, and a latent representation typically of size $768$ is calculated for each patch. Extracting features from $I_q$ and $I_g$ results in two vectors of size $[196, 768]$.}

{We then calculate the patch-wise similarity score, resulting in $196$.
Note that we can use patches of other granularities. In the extreme case, we can just use the CLS token of \texttt{ViT}, which gives $768$ latent space features of the whole image. But this may overlook some details. So we choose the most fine-grained patches available.}

{If we query the target model $m$ times, we obtain a similarity score vector of size $[m, 196]$ for all generated images. By applying our defined statistical function $f$, we aggregate the similarity scores from multiple generated images to produce a final similarity score vector of size $196$. This vector is then used as input for threshold-based, distribution-based, and classifier-based attack models.}

\section{More Details for Different Size of Auxiliary Dataset}

\begin{figure}[h]
    \centering
    \resizebox{0.45\textwidth}{!}{
    \includegraphics{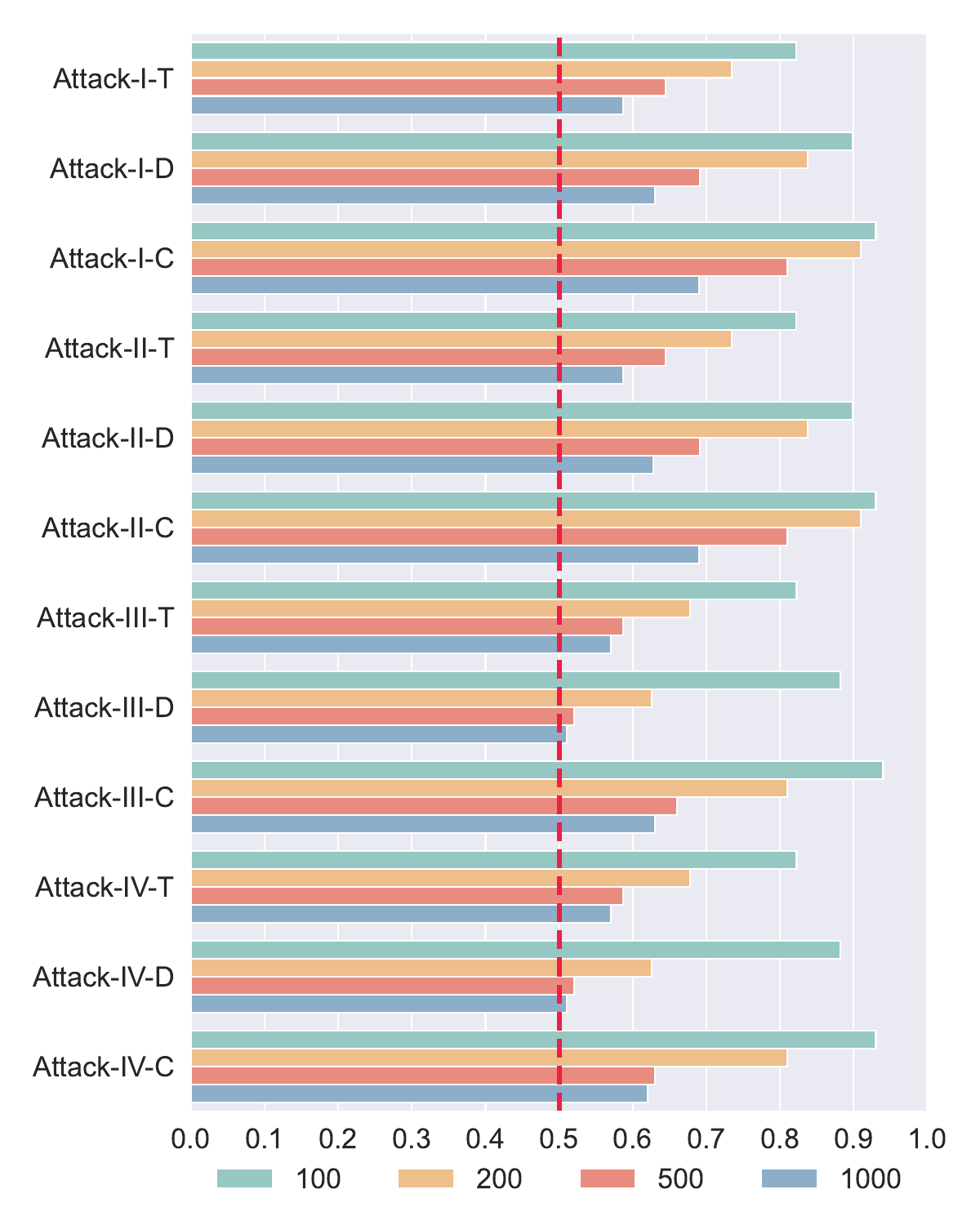}}
    \caption{Attack nomenclature and performance trends:`T' for threshold-based, `D' for distribution-based, and `C' for classifier-based attacks, with accuracy inversely related to training set size.}  
    \label{fig:size}
\end{figure}

{\autoref{fig:size} shows that in all attack scenarios, the attack performance decreases as the size of the auxiliary data increases. However, the classifier-based attack still maintains a ROC-AUC above $0.6$.}

\newpage
\onecolumn
\section{More Details for Comparing Five Different Image Encoders} \label{appendix:image_encoder}

To comprehensively analyze the influence of various image feature extractors on attack success rates, we evaluated the performance of five distinct image feature extractors across three types of attacks, within four attack scenarios obtained by the attacker, on three datasets. For each attack, we highlighted the optimal results of each evaluation metric across different image feature extractors. In~\autoref{appendix:encoder_classifier},~\autoref{appendix:encoder_threshold}, and~\autoref{appendix:encoder_distribution}, DeiT is the most stable image encoder and achieves the best attack performance.

\begin{table*}[h]
    \belowrulesep=0pt
    \aboverulesep=0pt
    \vspace{0.3cm}
    \caption{Comparative analysis of five different image encoders using \textit{classifier-based} attack across three datasets.}
    \label{appendix:encoder_classifier}
    \centering
    \LARGE
    \resizebox{0.97\textwidth}{!}{
    \begin{tabular}{c|c|ccccccccccccccc}
    \toprule[1.1pt]
       \multicolumn{2}{c|}{} & \multicolumn{3}{c}{\textbf{DETR}}  & \multicolumn{3}{c}{\textbf{BEiT}}   & \multicolumn{3}{c}{\textbf{EfficientFormer}}  & \multicolumn{3}{c}{\textbf{ViT}}  & \multicolumn{3}{c}{\textbf{DeiT}}  \\
       \cmidrule(lr){3-5} \cmidrule(lr){6-8} \cmidrule(lr){9-11} \cmidrule(lr){12-14} \cmidrule(lr){15-17}
        \multicolumn{2}{c|}{} & ASR & AUC & T@F=1\% & ASR & AUC & T@F=1\% &  ASR & AUC & T@F=1\% &  ASR & AUC & T@F=1\% &  ASR & AUC & T@F=1\% \\ \cmidrule[0.7pt](lr){1-17}
        \multirow{4}{*}{CelebA} & Attack-I & $0.66$ & $0.70$ & $0.10$ & $\bm{0.87}$ & $\bm{0.95}$ & $\bm{0.64}$ & $0.80$ & $0.87$ & $0.37$ & $0.81$ & $0.88$ & $0.26$ & $0.87$ & $0.93$ & $0.49$ \\
        & Attack-II & $0.67$ & $0.69$ & $0.09$ & $0.88$ & $0.94$ & $0.57$ & $0.82$ & $0.88$ & $0.38$ & $0.80$ & $0.88$ & $0.29$ & $\bm{0.88}$ & $\bm{0.94}$ & $\bm{0.61}$ \\
        & Attack-III & $0.67$ & $0.71$ & $0.07$ & $0.84$ & $0.91$ & $\bm{0.57}$ & $0.81$ & $0.87$ & $0.42$ & $0.79$ & $0.83$ & $0.40$ & $\bm{0.87}$ & $\bm{0.94}$ & $0.52$\\
        & Attack-IV & $0.67$ & $0.71$ & $0.10$ & $0.84$ & $0.91$ & $0.58$ & $0.78$ & $0.84$ & $0.44$ & $0.78$ & $0.83$ & $0.38$ & $\bm{0.88}$ & $\bm{0.93}$ & $\bm{0.60}$ \\
        \cmidrule[0.7pt](lr){1-17}
        \multirow{4}{*}{WIT} & Attack-I & $0.74$ & $0.79$ & $0.11$ & $0.70$ & $0.80$ & $0.30$ & $0.76$ & $0.81$ & $0.13$ & $0.77$ & $0.83$ & $0.06$ & $\bm{0.79}$ & $\bm{0.84}$ & $\bm{0.22}$ \\
        & Attack-II & $0.73$ & $0.77$ & $0.10$ & $0.69$ & $0.77$ & $\bm{0.27}$ & $0.71$ & $0.78$ & $0.11$ & $0.74$ & $0.80$ & $0.16$ & $\bm{0.78}$ & $\bm{0.85}$ & $0.15$ \\
        & Attack-III & $0.65$ & $0.72$ & $0.07$ & $0.71$ & $0.78$ & $0.17$ & $0.78$ & $0.82$ & $0.22$ & $0.78$ & $0.82$ & $0.21$ & $\bm{0.77}$ & $\bm{0.83}$ & $\bm{0.29}$ \\
        & Attack-IV & $0.64$ & $0.69$ & $0.08$ & $0.72$ & $0.77$ & $0.11$ & $0.76$ & $0.81$ & $0.16$ & $0.77$ & $0.82$ & $0.05$ & $\bm{0.75}$ & $\bm{0.83}$ & $\bm{0.25}$\\
        \cmidrule[0.7pt](lr){1-17}
        \multirow{4}{*}{MS COCO} & Attack-I & $0.72$ & $0.75$ & $0.17$ & $0.77$ & $0.84$ & $0.24$ & $0.78$ & $0.87$ & $0.20$ & $0.73$ & $0.82$ & $0.20$ & $\bm{0.85}$ & $\bm{0.93}$ & $\bm{0.61}$ \\
        & Attack-II & $0.75$ & $0.80$ & $0.06$ & $0.77$ & $0.85$ & $0.16$ & $0.81$ & $0.87$ & $0.35$ & $0.75$ & $0.83$ & $0.20$ & $\bm{0.85}$ & $\bm{0.92}$ & $\bm{0.56}$\\
        & Attack-III & $0.70$ & $0.78$ & $0.16$ & $0.78$ & $0.84$ & $\bm{0.44}$ & $0.78$ & $0.82$ & $0.28$ & $0.71$ & $0.80$ & $0.20$ & $\bm{0.83}$ & $\bm{0.89}$ & $0.30$\\
        & Attack-IV & $0.70$ & $0.76$ & $0.20$ & $\bm{0.80}$ & $\bm{0.83}$ & $\bm{0.40}$ & $0.76$ & $0.83$ & $0.27$ & $0.75$ & $0.82$ & $0.31$ & $0.69$ & $0.74$ & $0.16$\\
        \bottomrule[1.1pt]
    \end{tabular}}
\end{table*}

\begin{table*}[h]
    \belowrulesep=0pt
    \aboverulesep=0pt
    \vspace{0.3cm}
    \caption{Comparative analysis of five different image encoders using \textit{threshold-based} attack across three datasets.}
    \label{appendix:encoder_threshold}
    \centering
    \LARGE
    \resizebox{0.97\textwidth}{!}{
    \begin{tabular}{c|c|ccccccccccccccc}
    \toprule[1.1pt]
       \multicolumn{2}{c|}{} & \multicolumn{3}{c}{\textbf{DETR}}  & \multicolumn{3}{c}{\textbf{BEiT}}   & \multicolumn{3}{c}{\textbf{EfficientFormer}}  & \multicolumn{3}{c}{\textbf{ViT}}  & \multicolumn{3}{c}{\textbf{DeiT}}  \\
       \cmidrule[0.7pt](lr){3-5} \cmidrule[0.7pt](lr){6-8} \cmidrule[0.7pt](lr){9-11} \cmidrule[0.7pt](lr){12-14} \cmidrule[0.7pt](lr){15-17}
        \multicolumn{2}{c|}{} & ASR & AUC & T@F=1\% & ASR & AUC & T@F=1\% &  ASR & AUC & T@F=1\% &  ASR & AUC & T@F=1\% &  ASR & AUC & T@F=1\% \\ \cmidrule[0.7pt](lr){1-17}
        \multirow{4}{*}{CelebA} & Attack-I & $0.57$ & $0.64$ & $0.02$ & $\bm{0.79}$ & $\bm{0.86}$ & $0.41$ & $0.70$ & $0.76$ & $0.26$ & $0.73$ & $0.78$ & $0.01$ & $0.75$ & $0.82$ & $\bm{0.43}$ \\
        & Attack-II & $0.60$ & $0.64$ & $0.02$ & $\bm{0.77}$ & $\bm{0.86}$ & $0.41$ & $0.70$ & $0.76$ & $0.29$ & $0.73$ & $0.78$ & $0.18$ & $0.76$ & $0.82$ & $\bm{0.45}$\\
        & Attack-III & $0.57$ & $0.65$ & $0.01$ & $\bm{0.79}$ & $\bm{0.86}$ & $\bm{0.40}$ & $0.71$ & $0.76$ & $0.23$ & $0.67$ & $0.79$ &$ 0.12$& $0.75$ & $0.82$ & $0.20$ \\
        & Attack-IV & $0.59$ & $0.66$ & $0.05$ & $\bm{0.78}$ & $\bm{0.86}$ & $\bm{0.39}$ & $0.68$ & $0.76$ & $0.24$ & $0.69$ & $0.76$ & $0.13$ & $0.77$ & $0.84$ & $0.32$\\
        \cmidrule[0.7pt](lr){1-17}
        \multirow{4}{*}{WIT} & Attack-I & $0.59$ & $0.66$ & $0.07$ & $0.56$ & $0.62$ & $0.06$ & $0.66$ & $0.76$ & $0.12$ & $\bm{0.72}$ & $\bm{0.79}$ & $0.02$ & ${0.68}$ & ${0.77}$ & $\bm{0.17}$ \\
        & Attack-II & $0.57$ & $0.66$ & $\bm{0.14}$ & $0.57$ & $0.62$ & $0.05$ & $0.61$ & $\bm{0.70}$ & $0.03$ & $0.57$ & $0.61$ & $0.01$ & $\bm{0.62}$ & ${0.67}$ & $0.02$ \\
        & Attack-III & $0.60$ & $0.66$ & $0.02$ & $0.60$ & $0.61$ & $0.03$ & $\bm{0.66}$ & $\bm{0.71}$ & $0.09$ & $0.63$ & $\bm{0.71}$ & $0.03$ & $0.64$ & $0.69$ & $\bm{0.13}$ \\
        & Attack-IV & $0.57$ & $0.62$ & $0.01$ & $0.66$ & $0.73$ & $0.03$ & $0.69$ & $0.80$ & $0.13$ & $0.70$ & $\bm{0.82}$ & $0.02$ & $\bm{0.73}$ & $0.80$ & $\bm{0.20}$ \\
        \cmidrule[0.7pt](lr){1-17}
        \multirow{4}{*}{MS COCO} & Attack-I & $0.71$ & $0.73$ & $0.01$ & $0.71$ & $0.80$ & $0.14$ & $0.74$ & $0.82$ & $0.11$ & $0.68$ & $0.77$ & $\bm{0.17}$ & $\bm{0.79}$ & $\bm{0.84}$ & $0.05$ \\
        & Attack-II & $0.63$ & $0.72$ & $0.01$ & $0.75$ & $0.80$ & $0.15$ & $0.75$ & $0.82$ & $0.11$ & $0.70$ & $0.78$ & $\bm{0.18}$ & $\bm{0.79}$ & $\bm{0.85}$ & $0.06$ \\
        & Attack-III & $0.63$ & $0.72$ & $0.01$ & $0.69$ & $0.81$ & $\bm{0.23}$ & $\bm{0.76}$ & $0.82$ & $0.15$ & $0.68$ & $0.79$ & $0.01$ & $\bm{0.76}$ & $\bm{0.84}$ & $0.13$ \\
        & Attack-IV & $0.61$ & $0.72$ & $0.09$ & $0.71$ & $0.81$ & $\bm{0.24}$ & $\bm{0.78}$ & $\bm{0.82}$ & $0.17$ & $0.68$ & $0.79$ & $0.24$ & $0.70$ & $0.74$ & $0.01$ \\
        \bottomrule[1.1pt]
    \end{tabular}}
\end{table*}

\begin{table*}[h]
    \belowrulesep=0pt
    \aboverulesep=0pt
    \vspace{0.3cm}
    \caption{Comparative analysis of five different image encoders using \textit{distribution-based} attack across three datasets.}
    \label{appendix:encoder_distribution}
    \centering
    \LARGE
    \resizebox{0.97\textwidth}{!}{
    \begin{tabular}{c|c|ccccccccccccccc}
    \toprule[1.1pt]
       \multicolumn{2}{c|}{} & \multicolumn{3}{c}{\textbf{DETR}}  & \multicolumn{3}{c}{\textbf{BEiT}}   & \multicolumn{3}{c}{\textbf{EfficientFormer}}  & \multicolumn{3}{c}{\textbf{ViT}}  & \multicolumn{3}{c}{\textbf{DeiT}}  \\
       \cmidrule(lr){3-5} \cmidrule(lr){6-8} \cmidrule(lr){9-11} \cmidrule(lr){12-14} \cmidrule(lr){15-17}
        \multicolumn{2}{c|}{} & ASR & AUC & T@F=1\% & ASR & AUC & T@F=1\% &  ASR & AUC & T@F=1\% &  ASR & AUC & T@F=1\% &  ASR & AUC & T@F=1\% \\ \cmidrule[0.7pt](lr){1-17}
        \multirow{4}{*}{CelebA} & Attack-I & $0.62$ & $0.66$ & $0.03$ & $\bm{0.76}$ & $\bm{0.90}$ & $\bm{0.65}$ & $0.70$ & $0.83$ & $0.41$ & $0.74$ & $0.84$ & $0.40$ & $0.73$ & $\bm{0.90}$ & $0.61$ \\
        & Attack-II & $0.61$ & $0.66$ & $0.03$ & $\bm{0.79}$ & $\bm{0.90}$ & $\bm{0.66}$ & $0.71$ & $0.82$ & $0.32$ & $0.73$ & $0.85$ & $0.39$ & $0.74$ & $\bm{0.90}$ & $0.64$ \\
        & Attack-III & $0.56$ & $0.58$ & $0.01$ & $0.74$ & $0.85$ & $\bm{0.51}$ & $0.61$ & $0.71$ & $0.20$ & $0.64$ & $0.74$ & $0.13$ & $\bm{0.76}$ & $\bm{0.88}$ & $0.50$ \\
        & Attack-IV & $0.59$ & $0.61$ & $0.01$ & $0.72$ & $0.86$ & $\bm{0.61}$ & $0.61$ & $0.70$ & $0.21$ & $0.67$ & $0.73$ & $0.16$ & $\bm{0.77}$ & $\bm{0.88}$ & $0.52$ \\
        \cmidrule[0.7pt](lr){1-17}
        \multirow{4}{*}{WIT} & Attack-I &$0.66$& $0.72$ & $0.12$ & $\bm{0.70}$ & $0.83$ & $0.27$ & $\bm{0.70}$ & $0.82$ & $0.22$ & $\bm{0.70}$ & $\bm{0.85}$ & $0.30$ & $0.69$ & $0.84$ & $\bm{0.41}$\\
        & Attack-II & $0.58$ & $0.68$ & $0.07$ & ${0.70}$ & $0.81$ & $0.14$ & $0.66$ & $0.78$ & $0.19$ & $\bm{0.71}$ & $\bm{0.84}$ & $0.23$ & $0.68$& $\bm{0.84}$ & $\bm{0.34}$ \\
        & Attack-III & $0.57$ & $0.57$ & $0.01$ & $\bm{0.62}$ & $0.69$ & $0.15$ & $\bm{0.62}$ & $0.68$ & $0.20$ & $0.60$ & $0.67$ & $0.10$ & $0.61$ & $\bm{0.70}$ & $\bm{0.26}$\\
        & Attack-IV & $0.51$ & $0.55$ & $0.01$ & $0.61$ & $\bm{0.71}$ & $0.14$ & $0.60$ & $0.66$ & $\bm{0.23}$ & $0.56$ & $0.64$ & $0.11$ & $\bm{0.64}$ & $0.69$ & $0.09$ \\
        \cmidrule[0.7pt](lr){1-17}
        \multirow{4}{*}{MS COCO} & Attack-I & $0.62$ & $0.71$ & $0.19$ & $0.65$ & $0.80$ & $0.33$ & $0.67$ & $0.79$ & $0.14$ & $0.59$ & $0.73$ & $\bm{0.43}$ & $\bm{0.73}$ & $\bm{0.81}$ & $0.36$\\
        & Attack-II & $0.60$ & $0.71$ & $0.18$ & $0.64$ & $0.80$ & $0.32$ & $0.68$ & $0.80$ & $0.15$ & $0.61$ & $0.74$ & $\bm{0.41}$ & $\bm{0.72}$ & $\bm{0.81}$ & $0.36$ \\
        & Attack-III & $0.56$ & $0.57$ & $0.03$ & $0.62$ & $0.70$ & $0.09$ & $0.61$ & $0.63$ & $0.10$ & $0.61$ & $0.67$ & $0.06$ & $\bm{0.70}$ & $\bm{0.77}$ & $\bm{0.13}$ \\
        & Attack-IV & $0.55$ & $0.57$ & $0.02$ & $\bm{0.63}$ & $\bm{0.70}$ & $0.06$ & $0.59$ & $0.63$ & $\bm{0.12}$ & $0.62$ & $0.67$ & $0.07$ & $0.62$ & $0.67$ & $0.07$ \\
        \bottomrule[1.1pt]
    \end{tabular}}
\end{table*}

\onecolumn
\newpage
\section{More Experimental Results for Varying Fine-tuning Steps} \label{appendix:fine_steps}

\begin{figure*}[h]
    \centering
    \includegraphics[width=0.97\textwidth]{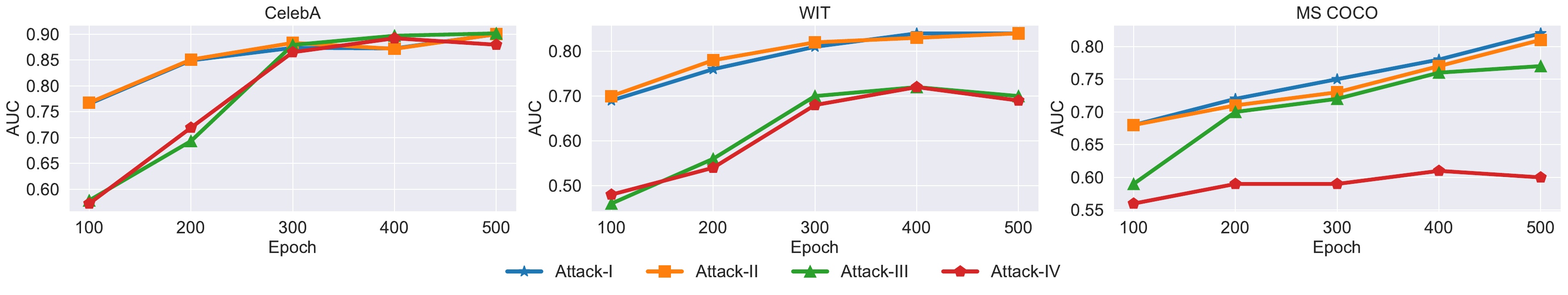}
    \caption{Correlation between increased fine-tuning steps and enhanced accuracy of \textit{distribution-based} attack.}
    \label{fig:distribution}
\end{figure*}

\begin{figure*}[h]
    \centering
    \includegraphics[width=0.97\textwidth]{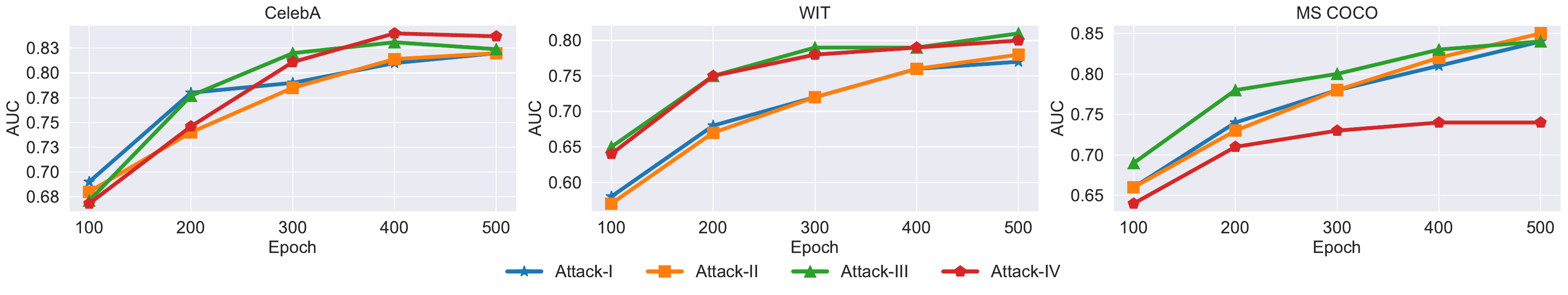}
    \caption{Increase in success rate of \textit{threshold-based} attack with more fine-tuning steps}
    \label{fig:thr}
\end{figure*}

In this part, we want to examine the impact of increasing fine-tuned steps on the outcomes of different types of attacks. The distribution-based attack results can be found in~\autoref{fig:distribution}, and the threshold-based attack is illustrated in~\autoref{fig:thr}. All these experiment results show that attack accuracy increases with more fine-tuning steps.

\section{More Experimental Results for Different Number of Inference Steps} \label{appendix:inference_step}

To evaluate how inference steps affect attack performance, we conducted experiments on the WIT and MS COCO datasets, with results detailed in~\autoref{tab:appendix_inference}. We highlighted the best attack results for each evaluation metric across different inference steps. The results indicate that the inference steps do not affect attack accuracy.

\vspace{1cm}

\begin{table}[ht]
    
    \centering
    \caption{Experiment results for more inference steps on MS COCO and WIT. The best attack result is marked in bold. }
    \label{tab:appendix_inference}
    \LARGE
    \resizebox{\textwidth}{!}{
    \begin{tabular}{ccccccccccccccccccccc}
    \toprule[1.1pt]
         \multirow{3}{*}{S} & \multicolumn{10}{c}{MS COCO} &  \multicolumn{10}{c}{WIT} \\ \cmidrule[0.7pt](lr){2-10} \cmidrule[0.7pt](lr){12-20}
         &\multicolumn{3}{c}{Threshold-based} & \multicolumn{3}{c}{Distribution-based} & \multicolumn{3}{c}{Classifier-based} & \multirow{2}{*}{FID} &\multicolumn{3}{c}{Threshold-based} & \multicolumn{3}{c}{Distribution-based} & \multicolumn{3}{c}{Classifier-based} & \multirow{2}{*}{FID} \\
         \cmidrule[0.7pt](lr){2-4} \cmidrule[0.7pt](lr){5-7} \cmidrule[0.7pt](lr){8-10} \cmidrule[0.7pt](lr){12-14} \cmidrule[0.7pt](lr){15-17} \cmidrule[0.7pt](lr){18-20}
         &ASR & AUC & T@F=1\% & ASR & AUC & T@F=1\% & ASR & AUC & T@F=1\% &  &ASR & AUC & T@F=1\% & ASR & AUC & T@F=1\% & ASR & AUC & T@F=1\%&  \\
         \midrule[0.7pt]
         $30$ & $0.76$ & $\bm{0.84}$ & $0.13$ & $0.70$ & $\bm{0.77}$ & $\bm{0.13}$ & $0.84$ & $0.90$  & $\bm{0.42}$  &  $8.49$ & $\bm{0.71}$ & $\bm{0.81}$ & $\bm{0.23}$ & $0.61$ & $0.70$ & $\bm{0.26}$ & $\bm{0.78}$ & $0.82$  & $0.29$  &  $6.73$\\
         $50$ & $0.74$ & $\bm{0.84}$ & $0.13$ & $0.69$ & $\bm{0.77}$ & $0.11$ & $0.84$ & $\bm{0.91}$  & $0.20$  &  $7.24$ &$\bm{0.71}$ & $0.80$ & $0.20$ & $0.62$ & $0.72$ & $0.25$ & $0.75$ & $0.82$  & $0.30$  &  $5.83$\\
         $100$ & $0.76$ & $\bm{0.84}$ & $0.15$ & $0.70$ & $0.76$ & $0.11$ & $\bm{0.85}$ & $0.90$  & $0.23$  &  $6.46$  &$\bm{0.71}$ & $0.79$ & $0.17$ & $\bm{0.65}$ & $\bm{0.74}$ & $0.09$ & $0.76$ & $\bm{0.83}$  & $0.32$  &  $5.58$ \\
         $200$ & $\bm{0.77}$ & $\bm{0.84}$ & $\bm{0.16}$ & $\bm{0.71}$ & $0.75$ & $0.11$ & $0.83$ &  $0.88$ &  $0.21$ & $6.46$ &$0.70$ & $0.79$ & $0.20$ & $0.62$ & $0.72$ & $0.14$ & $0.77$ &  $\bm{0.83}$ &  $\bm{0.33}$ & $5.56$\\
    \bottomrule[1.1pt]
    \end{tabular}}
\end{table}

\newpage

\end{document}